\documentclass{aa}

\usepackage{times}
\usepackage{graphicx}
\usepackage{xspace}
\usepackage{epsfig}
\usepackage{natbib}
\usepackage{rotating}

\begin{document}

\title{Further X-ray detections of Herbig stars}

\author{B. Stelzer\inst {1} \and J. Robrade\inst {2} \and J. H. M. M. Schmitt\inst {2} \and J. Bouvier\inst {3}}

\offprints{B. Stelzer}

\institute{INAF - Osservatorio Astronomico di Palermo,
  Piazza del Parlamento 1,
  I-90134 Palermo, Italy \\ \email{B. Stelzer, stelzer@astropa.unipa.it} \and
  Hamburger Sternwarte, 
  Gojenbergsweg 112,
  D-21029 Hamburg, Germany \and 
  Laboratoire d'Astrophysique,
  Observatoire de Grenoble,
  Universit\'e Joseph Fourier, B.P.\,53, 
  F-38041 Grenoble Cedex\,9, France}

\titlerunning{Further X-ray detections of Herbig stars}

\date{Received $<08-07-2008>$ / Accepted $<03-10-2008>$}

\abstract
{The interpretation of X-ray detections from Herbig Ae/Be stars is disputed as it is not clear
if these intermediate-mass pre-main sequence stars are able to drive a dynamo and ensuing phenomena
of magnetic activity. Alternative X-ray production mechanisms, 
related to stellar winds, star-disk magnetospheres, or unresolved late-type T Tauri star companions 
have been proposed. 
}
{The companion hypothesis can be tested by resolving Herbig stars in X-rays from their known visual 
secondaries. Furthermore, their global X-ray properties (such as detection rate, luminosity, temperature,
variability), may give clues to the emission mechanism by comparison to other types of
stars, e.g. similar-age but lower-mass T Tauri stars, similar-mass but more evolved
main-sequence A- and B-type stars, and with respect to model predictions.  
}
{In a series of papers we have been investigating high-resolution X-ray Chandra images of 
Herbig Ae/Be and main-sequence B-type stars where known close visual companions are spatially
separated from the primaries.
}
{Here we report on six as yet unpublished Chandra exposures from our X-ray survey of Herbig stars. 
The target list comprises six Herbig stars with known cool companions, and three further A/B-type stars
that are serendipitously in the Chandra field-of-view.  
In this sample we record a detection rate of $100$\,\%, i.e. all A/B-type stars display
X-ray emission at levels of $\log{(L_{\rm x}/L_{\rm bol})} \sim -5...-7$. 
The analysis of hardness ratios confirms that HAeBe's have hotter and/or more absorbed 
X-ray emitting plasma than more evolved B-type stars. 
}
{Radiative winds are ruled out as exclusive emission mechanism on basis of the high X-ray temperatures.
Confirming earlier results, 
the X-ray properties of Herbig Ae/Be stars are not vastly different from those of their late-type
companion stars (if such are known). The diagnostics provided by the presently available data
leave open if the hard X-ray emission of Herbig stars is due to young age or indicative of further
coronally active low-mass companion stars. In the latter case, our detection statistics imply a
high fraction of higher-order multiple systems among Herbig stars. 
}  


\maketitle

\section{Introduction}\label{sect:intro}

Theoretical considerations distinguish two major X-ray emission mechanisms 
for stars: Hot stars have strong winds in which shocks heat the plasma 
to a few $10^6$\,K, giving rise to soft X-ray emission.
Cool stars have dynamo-generated magnetic fields that, under the influence of
convection, provide a heating agent for the outer atmospheres, 
resulting in harder ($\geq 10$\,MK) X-ray emission.

For stars on the main-sequence (MS), there is a critical zone in between these two
regimes at spectral types mid-B and mid-A, 
in which stars have neither strong winds nor convective envelopes,
and thus are expected to be no X-ray emitters. Nevertheless, X-ray
detections of A- and B-type stars have been reported in several works; see 
in particular catalogs 
based on the spatially complete {\em ROSAT} All-Sky Survey 
\citep{Berghoefer96.1, Huensch98.1}. 
In absence of a physical explanation, the X-ray emission of these 
stars is generally attributed to unknown, unresolved late-type companions.
\citet{Stelzer03.1} and 
\citet{Stelzer06.2} have checked the companion hypothesis by resolving a sample of MS B-type
stars with {\em Chandra} from their close visual companions. The result was ambiguous, because
more than half of the B-type stars were detected with {\em Chandra} even after being
resolved from all known visual companions. However, the X-ray sources associated with the B-type
stars do not show significantly different properties from those coincident with their cooler 
companion stars. This is fully consistent with the idea that the X-rays originate from 
even closer spectroscopic companions.

The case is more complex for intermediate-mass stars on the pre-main sequence (pre-MS), the so-called
Herbig Ae/Be stars. 
The relatively low-sensitivity instruments onboard {\em ROSAT}
and {\em ASCA} achieved detection fractions between $30-50$\,\% 
\citep{Zinnecker94.1, Hamaguchi05.1}. 
Contrary to MS A- and B-type stars, HAeBe stars are surrounded
by circumstellar material in the form of disks and envelopes, remnants of the star forming process. 
Lately, it was recognized that magnetic coupling between star and accretion disk leads to important modifications
of the X-ray properties of {\it low-mass} pre-MS stars: X-ray emission seems
to be suppressed in accreting T Tauri stars \citep{Preibisch05.1, Telleschi07.1}, 
magnetic reconnection of the star-disk field has been held responsible
for flares \citep{Favata05.1}, and an excess of soft X-rays in some objects was ascribed 
to emission from either accretion shocks \citep{Kastner02.1, Stelzer04.3} 
or to shocks in outflows \citep{Guedel07.1}. 
These mechanisms could play a role also in HAeBe stars that are still in their accretion phase.
Indeed, unusually soft X-ray emission was observed for HD\,163296 and ascribed to emission from
an accretion shock similar to the case of the lower-mass T~Tauri stars \citep{Swartz05.1}. 
A possible link between X-ray activity of HAeBe stars and the presence of outflows, 
typical for stars in the accretion phase, was suggested by \citet{Hamaguchi05.1}. 
Indications for the decline of the
X-ray luminosity with age in a sample of HAeBe stars observed with
{\em ASCA} have been interpreted in terms of star-disk magnetic interactions,
similar as proposed for protostars \citep{Hamaguchi05.1}.
Last but not least, X-ray flares from HAeBe stars were observed with {\em ASCA} and {\em XMM-Newton}
\citep{Hamaguchi00.1, Giardino04.1}, 
but the association with the HAeBe stars was not unambiguous. 

HAeBe stars are progenitors of MS A- and B-type stars. Since a fraction of $\sim 5$\,\% of 
intermediate-mass stars on the MS is magnetic, the Ap/Bp stars, a similar or higher fraction of 
HAeBe stars is expected to harbor magnetic fields as remnants from the star forming process. 
While initial attempts to measure those fields remained unsuccessful, with the advent of 
new spectropolarimetric instrumentation magnetic field detections of HAeBe stars are starting
to accumulate. 
At present fields have been positively measured for eight HAeBe stars, corresponding to a 
fraction of $7$\,\% of the samples studied, in accordance with extrapolations of the magnetic field
incidence of MS stars. 
In addition, and in contrast to the A- and B-type stars on the MS, HAeBe stars may also possess a shear dynamo
that is supplied by the rotational energy of the star, and was shown to sustain magnetic fields
in the initial phase (a few Myrs) of the life of an intermediate-mass star \citep{Tout95.1}.
However, from the position on fully radiative evolutionary tracks of the HAeBe stars
with field detections and the large-scale ordered structure
of their fields \cite{Alecian08.1} conclude that these fields are primordial rather than dynamo generated.
The efficiency of such global fields in generating X-rays is unclear. 

Another possible mechanism for X-ray production in HAeBe stars regards a picture in which 
magnetic fields influence the wind geometry and dynamics by channeling the wind. This scenario, 
referred to as magnetically confined wind shock (MCWS) model, was originally developed
to explain the X-ray emission from the Ap star IQ\,Aur \citet{Babel97.1}. It was also successfully
applied to the O-type star $\Theta$\,Ori\,C \citep{Babel97.2, Gagne05.1} and it could be the cause for the
strong and variable X-ray emission of some hot stars in Orion \citep{Stelzer05.1}. 
Its application to HAeBe stars depends on whether wind velocities and X-ray temperatures 
can be reconciled. 
This seems to be dubious due to the generally slow HAeBe winds ($\leq 600$\,km/s) 
and high X-ray temperatures ($>1$\,keV) measured e.g. by \cite{Skinner04.1} and by \cite{Stelzer06.3}.  
Nevertheless, a MCWS is considered to be the most likely explanation for the X-ray emission from the
HAeBe star AB\,Aur \citep{Telleschi07.2}. AB\,Aur is the first HAeBe star for which
a high-resolution X-ray spectrum was obtained. Its excess of soft emission, diagnosed by the high O\,VII line flux,
is similar to that of cTTS \citep{Guedel07.4} but, contrary to those latter ones, not accompanied by
high densities. Analysis of the He-like triplets indicates that the X-rays from AB\,Aur originate at a substantial
distance from the stellar surface in the wind region. A direct relation between the 
X-ray source and the wind of AB\,Aur is also supported by the variability of the X-ray lightcurve which 
was observed to be periodic with the same time-scale found in UV lines formed in the wind.

Despite the abundant speculations about the possible mechanism, 
a convincing and unique explanation for the X-ray emission of HAeBe stars has not been identified. 
Similar to the B- and A-type stars on the MS, T Tauri like 
companions could be the cause for the observed X-ray emission from HAeBe stars. 
The high X-ray luminosities of many HAeBe stars ($\log{L_{\rm x}}\,{\rm [erg/s]} \sim 30...31$),
have often been cited against the companion hypothesis, but \cite{Skinner04.1} 
showed that their range of X-ray luminosities is compatible with the typical emission 
level of a late-type pre-MS star. 
The binary fraction of HAeBe stars has been examined in high angular resolution
studies \citep{Leinert97.1, Pirzkal97.1}, and resulted in an
excess of binaries with respect to MS stars. The brightness ratios of the HAeBe 
binaries suggests that most companions are of significantly lower mass than the
primaries, i.e. they are T Tauri stars, and must be strong X-ray emitters by their
nature. 

{\em Chandra} is the only satellite that provides a sub-arcsecond spatial resolution in X-rays,
which is reasonably close to infrared (IR) imaging (adaptive optics) observations. 
This implies that the majority of known visual companions,
those at separations larger than $\sim 1^{\prime\prime}$, can be resolved for the first time in X-rays. 
Here we extend our previous archival {\em Chandra} study on HAeBe stars 
with the aim to investigate the possibility that the X-ray emission arises
from known T Tauri star companions.
In Sect.~\ref{sect:sample} we introduce the new targets 
observed with {\em Chandra}. The near-IR imaging survey from which the {\em Chandra} sample
was selected is described in Sect.~\ref{sect:ao}. The X-ray observations and the data analysis are 
presented in Sect.~\ref{sect:observations}, and results are given in   
Sect.~\ref{sect:results}. In Sect.~\ref{sect:discussion} we discuss the new detections
in the context of previous results, and Sect.~\ref{sect:conclusions} contains concluding remarks. 
Information on individual stars is found in the Appendix~\ref{sect:indiv}.

\section{The Sample}\label{sect:sample}

We based our target selection on the catalog of HAeBe stars and candidates by
\citet{The94.1}, and searched the literature for reports on close binaries among them.  
In addition we use the results of a recent high spatial resolution imaging 
survey for binaries among HAeBe stars that identified faint IR sources with separations in the range 
$0.1-8^{\prime\prime}$ near HAeBe stars (see Sect.~\ref{sect:ao}). 
From this analysis it results that $22$ stars of Table~1 from \citet{The94.1}
have close companions identified in IR adaptive optics images that are resolvable
with {\em Chandra} (separation $> 1^{\prime\prime}$).                            

{\em Chandra} observations for four of these have already been presented 
(V892\,Tau, HD\,141569, MWC\,863, and V380\,Ori; 
see discussion by \citet{Stelzer06.3}). 
We selected another five stars from the remaining list of $18$ for having 
well-constrained optical parameters, no evidence for being spectroscopic binaries (SBs), and 
yielding a good constraint on the X-ray emission (i.e. high bolometric luminosity or small distance). 
One of the {\em Chandra} fields comprises three Be stars (Obs-ID\,6399). 
This observation pointed towards the NGC\,7129 reflection nebula that is illuminated by three 
early B-type stars: the original target for our {\em Chandra} study, V373\,Cep (alias LkH$\alpha$\,234), the
Herbig Be star BD+65\,1637 \cite[also in the HAeBe star list of ][]{The94.1} and BD+65\,1638. 
BD+65\,1637 was observed within the IR imaging survey described in Sect.~\ref{sect:ao} but no companion
was detected. BD+65\,1638 is also not known to be binary. 
We include these two stars in our sample to enhance the data base
for X-ray observations of intermediate-mass stars.

We add to this sample a recent {\em Chandra} observation of HR\,5999, that fullfills our selection criteria:
HR\,5999 is a HAeBe star with a close visual T Tauri companion resolvable with 
{\em Chandra} \citep{Stecklum95.1}. Possible indications for a closer ($0.17$\,AU) low-mass companion 
were reported from extensive radial velocity monitoring \citep{Tjin89.1}.  
Another A-type star in the {\em Chandra} field, HR\,6000, 
shares common proper motion with HR\,5999 from which it is 
separated by $45^{\prime\prime}$. HR\,6000 is a chemically peculiar star and not known to have companion 
stars. It adds to the existing data base of X-ray emission from A-type stars. 

\begin{figure}
\begin{center}
\resizebox{9cm}{!}{\includegraphics{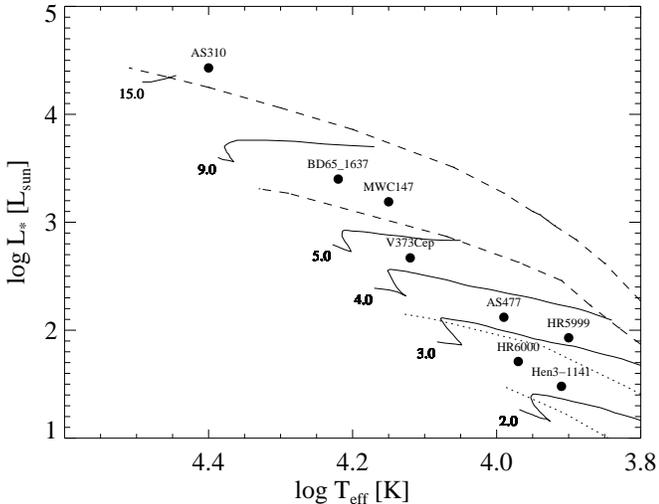}}
\caption{HR diagram for HAeBe stars observed with {\em Chandra}; BD\,+65\,1638 is not shown because its 
stellar parameters are unknown. 
Superposed on the data are evolutionary models by \protect\citet{Palla99.1}: solid lines are tracks labeled
by the corresponding mass in solar units, dotted lines are $1$ and $5$\,Myr isochrones and dashed lines
are the birthlines for accretion rates of $\dot{M} = 10^{-4}\,{\rm M_\odot/yr}$ and
$10^{-5}\,{\rm M_\odot/yr}$, respectively.}
\label{fig:hrd}
\end{center}
\end{figure}

The stellar parameters for the new {\em Chandra} sample are given in Table~\ref{tab:targets}. 
This sample spans a range of about $3$ dex
in bolometric luminosity and from $\sim 8000 - 25000$\,K in effective temperature, as 
illustrated in Fig.~\ref{fig:hrd} which represents the HR diagram for our targets on the 
evolutionary models by \citet{Palla99.1}. 
%
%
\begin{table*}
\begin{center}
\caption{Stellar parameters for the sample observed with {\em Chandra}. The citations in the last column refer to cols.~$4-10$.}
\label{tab:targets}
\begin{tabular}{lllcclrrrrc} \\ \hline
The No. & Name & Other Name     & SB? & VB ?    & SpT & d    & $\log{\frac{L_{\rm bol}}{L_\odot}}$ & $\log{T_{\rm eff}}$ & $A_{\rm V}$ & Ref \\
        &      &                &     &         &     & [pc] &                                     & [K]                 & [mag] & \\
\hline
   42  &  MWC147      & HD259431    &   N &  Y  & B6     & $ 800$ & $3.19 $ & $4.15$ &  $1.2$ & (1,2,3,4,5,5,5) \\
   69  &  Hen3-1141   & HD144432    &   N &  Y  & A7     & $ 145$ & $>1.48$ & $3.91$ &  $0.6$ & (1,6,7,8,4,4,4) \\
   70  &  HR\,5999    & HD144668    &   Y?&  Y  & A7     & $ 150$ & $1.94 $ & $3.90$ &  $0.5$ & (9,2,10,11,4,4,4) \\
   79  &  AS310       &             &     &  Y  & B0     & $2500$ & $4.43 $ & $4.40$ &  $4.1$ & (-,12,7,7,5,5,5)   \\
   98  &  BD+65\,1637 & V361\,Cep   &     &     & B3     & $1000$ & $3.40 $ & $4.22$ &  $1.8$ & (-,-,13,14,5,5,5) \\
   99  &  V373Cep     & LkHa234     &     &  Y  & B5/7   & $1000$ & $2.67 $ & $4.12$ &  $3.1$ & (-,15,15,7,5,5,5) \\
  100  &  AS477       & BD+46\,3471 &   N &  Y  & A0     & $ 900$ & $2.12 $ & $3.99$ &  $0.3$ & (1,16,7,7,5,5,5) \\
       &  BD+65\,1638 &             &     &     & B2     & $1000$ &         &        &        & (-,-,13,14,-,-,-)\\
       &  HR\,6000    & HD144667    &     &     & A1.5   & $ 150$ & $1.71 $ & $3.97$ &  $0.1$ & (-,-,17,11,17,17,17) \\
\hline
\multicolumn{11}{l}{(1) - \protect\cite{Corporon99.1}, (2) - Thomas et al., in prep., (3) - \protect\cite{Mottram07.1}, (4) - \protect\cite{vandenAncker98.1},} \\
\multicolumn{11}{l}{(5) - \protect\cite{Hernandez04.1}, (6) - \protect\cite{Dommanget94.1}, (7) - \protect\cite{Maheswar02.1}, (8) - \protect\cite{Perez04.1}, (9) - \protect\cite{Tjin89.1},} \\
\multicolumn{11}{l}{(10) - \protect\cite{Stecklum95.1}, (11) - \protect\cite{Hughes93.1}, (12) - \protect\cite{Ageorges97.1}, (13) - \protect\cite{Wang07.1},} \\
\multicolumn{11}{l}{(14) - \protect\cite{Hillenbrand92.1}, (15) - \protect\cite{Leinert97.1}, (16) - \protect\cite{Pirzkal97.1}, (17) - \protect\cite{GarciaLopez06.1}.} \\ 
\end{tabular}
\end{center}
\end{table*}

\section{High-resolution near-IR imaging}\label{sect:ao}

Near-IR imaging for 
the {\em Chandra} sample presented in Sect.~\ref{sect:sample} 
had been performed between 1993 and 1996 as part of a high angular
resolution survey of visual binaries among HAeBe stars \citep{Bouvier01.1}. 
MWC\,147, Hen\,3-1141 and HR\,5999 were observed with the
ADONIS adaptive optics (AO) system at ESO \citep{Beuzit97.1} equipped with the SHARP
IR camera \citep{Hofmann95.1}. AS\,310, V373\,Cep, BD+65\,1637 and AS\,477 were
observed with the PUE'O AO system at CFHT \citep{Rigaut98.1}
equipped with the Monica IR camera \citep{Nadeau94.1}.  Both
systems deliver nearly diffraction-limited images at $K$ with a FWHM of
$\simeq$ 0.1$\arcsec$ and a pixel sampling of $0.0344\,\arcsec$ at CFHT
and $0.051\,\arcsec$ at ESO. 

Exposure times ranged typically between $60$ and $180$\,s in each $JHK$
filter. To produce the final images, individual exposures were first
dark and flat corrected, then registered to subpixel accuracy and
eventually added. Astrometric standards from \cite{vanDessel93.1} 
were observed during each near-IR run in order to
derive the detector's orientation and plate scale. The accuracy of the
astrometric calibration is typically of order of a fraction of a
degree for the position angle and of less than $10^{-4}$\,arcsec per
pixel for the plate scale. 

Results from this AO survey will be published by Thomas et al., in prep.
We anticipate here the detection of faint companions to our {\em Chandra} sample.
The list of {\em Chandra} targets with all known visual companions is summarized in Table~\ref{tab:obslog}. 
All binary separations and position angles
reported in Table~\ref{tab:obslog} were measured on the near-IR images. Aperture and/or PSF
photometry was performed on the resolved multiple systems to measure
flux ratios between the components. The photometric accuracy is of
typically $0.05$\,mag at near-IR wavelengths. 

The measured flux ratios for the newly identified
companions are large indicating that those objects have much lower mass than the primary
Herbig stars if they form coeval systems. Thus they are probably T\,Tauri stars and it is
plausible to expect strong X-ray emission from them.
%
%
\begin{table*}
\begin{center}
\caption{Target list and observing log. Separation, position angle, $K$ band flux ratio, and identifier flag for all known companions (see Sect.~\ref{sect:ao}). The {\em Chandra}/ACIS observing log comprises observation ID and exposure time. Obs-IDs 6397...6401 were obtained with ACIS-S, and Obs-ID 8901 with ACIS-I. The last two columns represent the boresight correction applied to the X-ray data after cross-correlation with the 2\,MASS catalog (see text in Sect.~\ref{sect:observations})} 
\label{tab:obslog}
\begin{tabular}{lrrrrrccrrr}
\noalign{\smallskip} \hline \noalign{\smallskip} 
\multicolumn{3}{c}{Primaries} & \multicolumn{4}{c}{Companions} & \multicolumn{4}{c}{ACIS observations} \\
Designation & \multicolumn{2}{c}{Position}                                               & \multicolumn{1}{c}{Sep}                   & \multicolumn{1}{c}{PA}         & $\Delta K$ & Component & ObsID & Expo & \multicolumn{2}{c}{Astrometric correction$^{(d)}$}                    \\
            & \multicolumn{1}{c}{$\alpha_{2000}$} & \multicolumn{1}{c}{$\delta_{2000}$}  & \multicolumn{1}{c}{[$^{\prime\prime}$]}   & \multicolumn{1}{c}{[$^\circ$]} & [mag]      &           &       & [s]  & $\Delta\alpha$[$^{\prime\prime}$] & $\Delta\delta$[$^{\prime\prime}$] \\
\noalign{\smallskip} \hline \noalign{\smallskip} 

MWC147            & 06:33:05.19 & $+$10:19:20.0 & 3.11   & 344.4       & $5.67$ & B & 6397 &  9342.6 & $+0.03$ & $-0.09$ \\
\noalign{\smallskip} \hline \noalign{\smallskip}
Hen\,3-1141       & 16:06:57.96 & $-$27:43:09.8 & 1.45   &   3.0       & $2.55$ & B & 6398 & 12789.6 & & \\
\noalign{\smallskip} \hline \noalign{\smallskip}
AS\,310           & 18:33:21.21 & $-$04:58:06.7 & 1.31   &  78.3       & $3.69$ & B & 6399 & 38198.1 & $-0.26$ & $+0.08$ \\
                  &             &               & 2.22   & 240.5       & $3.42$ & C &      &         & & \\ 
                  &             &               & 3.74   & 233.4       & $3.70$ & D &      &         & & \\ 
                  &             &               & 4.34   & 122.0       & $1.06$ & E &      &         & & \\ 
                  &             &               & 4.88   &   0.9       & $2.70$ & F &      &         & & \\ 
                  &             &               & 5.00   &   6.1       & $4.65$ & G &      &         & & \\ 
\noalign{\smallskip} \hline \noalign{\smallskip} 
V373\,Cep         & 21:43:06.68 & $+$66:06:54.6 & 1.87   &  96.6       & $5.50$ & B & 6400 & 22662.9 & $-0.27$ & $+0.10$ \\
BD+65\,1637       & 21:42:50.21 & $+66$:06:32.2 & $-$    &  $-$        & $$     &$-$&      &         & & \\
BD+65\,1638       & 21:42:58.80 & $+66$:06:10.0 & $-$    &  $-$        & $$     &$-$&      &         & & \\
\noalign{\smallskip} \hline \noalign{\smallskip}
AS\,477           & 21:52:34.10 & $+$47:13:43.6 & 1.31   & 308.7       & $5.01$ & B & 6401 & 26571.2 & $+0.21$ & $+0.07$ \\
                  &             &               & 4.67   &  40.0       & $4.48$ & C &      &         & & \\ 
                  &             &               & 5.82   & 205.7       & $5.21$ & D &      &         & & \\ 
                  &             &               & 6.01   & 199.9       & $$     & E &      &         & & \\ 
\noalign{\smallskip} \hline \noalign{\smallskip}
HR\,5999          & 16:08:34.29 & $-$39:06:18.3 & 1.46   & 109.7       & $$     & B & 8901 & 9899.3  & $+0.15$ & $+0.07$ \\
HR\,6000          & 16:08:34.56 & $-$39:05:34.3 & $-$    & $-$         & $$     &$-$&      &         & & \\
\noalign{\smallskip} \hline \noalign{\smallskip}
\multicolumn{11}{l}{$^{(d)}$ - No boresight correction could be performed for Obs-ID\,6398 because of the absence of any 2\,MASS identification in the {\em Chandra} image.} \\
\end{tabular}
\end{center}
\end{table*}

\section{X-ray observations and data analysis}\label{sect:observations}

All observations were performed with {\em Chandra}'s ACIS-S array
with the exception of Obs-ID\,8901 (HR\,5999) that was carried out with ACIS-I.
The data analysis was performed with the CIAO software 
package\footnote{CIAO is made available by the CXC and can be downloaded 
from \\ http://cxc.harvard.edu/ciao/download/} version 3.4. 
We started our analysis with the level\,1 events file provided by the
{\em Chandra} X-ray Center (CXC). 
Correction for the charge transfer inefficiency (CTI) had been applied during
standard pipeline processing at the CXC. 
In the process of converting the level\,1 events file to a level\,2 events file
for each of the observations we performed the following steps: 
We removed the pixel randomization which is automatically applied by the CXC pipeline
in order to optimize the spatial resolution. 
We filtered the events file for event grades
(retaining the standard grades $0$, $2$, $3$, $4$, and $6$), 
and applied the standard good time interval file. 
Events flagged as cosmic rays were not removed in our analysis. In principle, such events 
can lead to the detection of spurious sources. However, if identified on the position of a
bright X-ray source, the flag is often erroneous (as a result of the event pattern used for 
the identification of cosmic rays). 

For our science goal of detecting and separating the HAeBe stars and their close companions, 
source detection was restricted to a $50 \times 50$ pixels wide image 
(1\,pixel $= 0.25^{\prime\prime}$) centered on the position of each of the nine primary stars. 
Source detection was carried out with the {\sc wavdetect} algorithm \citep{Freeman02.1}.
This algorithm correlates the data with a mexican hat function
to search for deviations from the background. The {\sc wavdetect} 
mechanism is well suited for separating closely spaced point sources.  
We used wavelet scales between $1$ and $8$ in steps of $\sqrt{2}$. 
Generally, the detection significance was set to $10^{-6}$ to avoid spurious detections.
This threshold was lowered to $10^{-4}$ for
the stars with the closest companions, Hen\,3-1141, AS\,477, and HR\,5999.

To ensure high accuracy of the X-ray positions the X-ray data were cross-correlated
with 2\,MASS point sources. To this end, 
source detection was performed on a more coarsely binned ($0.5^{\prime\prime}$/pixel) and larger 
($5^\prime \times 5^\prime$) image. 
This area comprises the whole primary chip, and some part of the adjacent ACIS chips if they 
were turned on. 
For each field the positions of all detected X-ray sources were cross-correlated 
with the 2\,MASS Point Source Catalog \citep{Cutri03.1}. Subsequently, the X-ray coordinates were
shifted by the detected offsets that are summarized in the last two columns of Table~\ref{tab:obslog}. 
In Obs-ID\,6399, after the small boresight correction, the SIMBAD position of the primary 
AS\,310 is $1.1^{\prime\prime}$ south
of an X-ray source that coincides with the brightest 2\,MASS object in the field.
The $JHK$ colors of this object are consistent with an early-type star. A chance coincidence
of another near-IR and X-ray bright object ($K=9.9$\,mag) 
so close to AS\,310, while the star itself remains
undetected both in 2\,MASS and in the {\em Chandra} image, seems rather unlikely. Therefore, we assume
that the 2\,MASS object represents AS\,310, 
and we computed the positions of its companion candidates with respect to the 2\,MASS position. 

\begin{sidewaystable*}\begin{center}
\caption{X-ray parameters of all components in the sample.}
\label{tab:xrayparams_lx}
\begin{tabular}{lccrrrrrrrrrrrr}\hline
Designation & Opt/IR & X-ray Iden. & Offax       & $\Delta_{\rm xo}$                               & S/N & Counts$^*$ & \multicolumn{1}{c}{HR\,1} & \multicolumn{1}{c}{HR\,2} & PSF frac. & $\log{L_{\rm x}^*}$ & $\log{(L_{\rm x}^*/L_{\rm bol})}$ & $P_{\rm KS}$ \\
            &        &             & [$^\prime$] & \multicolumn{1}{c}{[$^{\prime\prime}$]}         &       &            &                           &                           & [\%]      & [erg/s]             &                                   &              \\ \hline
MWC\,147    & A & $\surd$  & $  0.12 $ & $   0.27 $ & $1030.4 $ & $  178.8 \pm   14.4$ & $  0.54 \pm   0.09$ & $  0.01 \pm   0.13$ & $ 0.90$ & $  31.2$ & $  -5.6$ & $ 0.07$ \\
MWC\,147    & B & $\surd$  & $  0.14 $ & $   0.10 $ & $ 200.9 $ & $   51.7 \pm    8.2$ & $  0.12 \pm   0.20$ & $ -0.03 \pm   0.28$ & $ 0.90$ & $  30.6$ & ...      & $ 0.67$ \\
HEN\,3-1141 & A & $\surd$  & $  0.12 $ & $   0.38 $ & $  14.2 $ & $   37.4 \pm    7.2$ & $ -0.46 \pm   0.19$ & $ -0.54 \pm   0.41$ & $ 0.64$ & $  28.9$ & $  -6.2$ & $ 0.30$ \\
HEN\,3-1141 & B & $\surd$  & $  0.14 $ & $   0.46 $ & $ 635.3 $ & $ 1743.3 \pm   42.8$ & $  0.17 \pm   0.03$ & $  0.03 \pm   0.04$ & $ 0.92$ & $  30.4$ & ...      & $ 0.18$ \\
AS\,310     & A & $\surd$  & $  0.19 $ & $   0.07 $ & $  31.5 $ & $   17.4 \pm    5.3$ & $  0.26 \pm   0.32$ & $ -0.50 \pm   0.43$ & $ 0.90$ & $  30.9$ & $  -7.1$ & $ 0.54$ \\
AS\,310     & B & $-$      & $  0.18 $ & $-$        & $  11.7 $ & $    6.4 \pm    3.7$ & $-$                 & $-$                 & $ 0.90$ & $  30.5$ & ...      & $-$ \\
AS\,310     & C & $-$      & $  0.21 $ & $-$        & $   5.4 $ & $    2.5 \pm    2.8$ & $-$                 & $-$                 & $ 0.90$ & $  30.1$ & ...      & $-$ \\
AS\,310     & D & $-$      & $  0.24 $ & $-$        & $  11.0 $ & $    5.5 \pm    3.5$ & $-$                 & $-$                 & $ 0.90$ & $  30.4$ & ...      & $-$ \\
AS\,310     & E & $-$      & $  0.22 $ & $-$        & $-$       & $ <     3.0$         & $-$                 & $-$                 & $ 0.90$ & $< 30.1$ & ...      & $-$ \\
AS\,310     & F & $-$      & $  0.11 $ & $-$        & $-$       & $ <     3.0$         & $-$                 & $-$                 & $ 0.90$ & $< 30.1$ & ...      & $-$ \\
AS\,310     & G & $-$      & $  0.11 $ & $-$        & $-$       & $ <     3.0$         & $-$                 & $-$                 & $ 0.90$ & $< 30.1$ & ...      & $-$ \\
V373\,CEP   & A & $-$      & $  0.11 $ & $-$        & $   7.8 $ & $    3.5 \pm    3.1$ & $-$                 & $-$                 & $ 0.90$ & $  29.5$ & $-6.7$   & $-$ \\
V373\,CEP   & B & $\surd$  & $  0.08 $ & $   0.43 $ & $  30.5 $ & $   12.6 \pm    4.7$ & $  1.00 \pm   0.00$ & $  1.00 \pm   0.00$ & $ 0.90$ & $  30.1$ & ...      & $ 0.26$ \\
BD65\,1637  & A & $\surd$  & $  1.80 $ & $   0.14 $ & $ 181.1 $ & $   73.6 \pm    9.6$ & $  0.49 \pm   0.14$ & $ -0.25 \pm   0.19$ & $ 0.90$ & $  30.7$ & $  -6.3$ & $ 0.16$ \\
BD65\,1638  & A & $\surd$  & $  1.14 $ & $   0.17 $ & $ 189.5 $ & $  196.0 \pm   15.0$ & $  0.57 \pm   0.08$ & $  0.12 \pm   0.11$ & $ 0.84$ & $  30.8$ & $...$    & $ 0.65$ \\
BD65\,1638  & B & $\surd$  & $  1.15 $ & $   0.03 $ & $ 283.7 $ & $  282.0 \pm   17.8$ & $  0.46 \pm   0.07$ & $ -0.09 \pm   0.10$ & $ 0.84$ & $  30.9$ & ...      & $ 0.14$ \\
AS\,477     & A & $\surd$  & $  0.12 $ & $   0.22 $ & $  51.7 $ & $   20.6 \pm    5.6$ & $  0.05 \pm   0.33$ & $  0.00 \pm   0.49$ & $ 0.84$ & $  29.7$ & $  -6.0$ & $ 0.43$ \\
AS\,477     & B & $\surd$  & $  0.15 $ & $   0.17 $ & $ 216.6 $ & $   83.6 \pm   10.2$ & $  0.36 \pm   0.14$ & $  0.00 \pm   0.19$ & $ 0.84$ & $  30.3$ & ...      & $ 0.67$ \\
AS\,477     & C & $\surd$  & $  0.15 $ & $   0.19 $ & $ 293.2 $ & $  183.4 \pm   14.6$ & $  0.36 \pm   0.10$ & $ -0.13 \pm   0.13$ & $ 0.90$ & $  30.7$ & ...      & $ 0.03$ \\
AS\,477     & D & $\surd$  & $  0.13 $ & $   0.46 $ & $  25.0 $ & $    7.7 \pm    3.9$ & $  1.00 \pm   0.00$ & $  0.14 \pm   0.59$ & $ 0.90$ & $  29.3$ & ...      & $ 0.89$ \\
AS\,477     & E & $-$      & $  0.12 $ & $-$        & $  18.4 $ & $    5.7 \pm    3.5$ & $-$                 & $-$                 & $ 0.90$ & $  29.1$ & ...      & $ 0.00$ \\
HR\,5999    & A & $-$      & $  0.49 $ & $-$        & $   7.5 $ & $    7.1 \pm    3.8$ & $-$                 & $-$                 & $ 0.62$ & $  28.3$ & $  -7.2$ & $-$ \\
HR\,5999    & B & $\surd$  & $  0.49 $ & $   0.23 $ & $ 847.2 $ & $  688.2 \pm   27.2$ & $  0.41 \pm   0.07$ & $ -0.06 \pm   0.09$ & $ 0.91$ & $  30.1$ & ...      & $ 0.57$ \\
HR\,6000    & A & $\surd$  & $  0.27 $ & $   0.26 $ & $ 554.5 $ & $  316.4 \pm   18.8$ & $  0.10 \pm   0.08$ & $ -0.38 \pm   0.11$ & $ 0.90$ & $  29.7$ & $  -5.6$ & $(D)$ \\
\hline
\multicolumn{13}{l}{$^*$ in the $0.2-8$\,keV passband;} \\
\multicolumn{13}{l}{$L_{\rm x}$ refers to the distance given in Table~\ref{tab:targets} and has been corrected for the PSF fraction of counts in the extraction radius given in col.\,10.} \\
\multicolumn{13}{l}{$(D)$ Periodic variability is due to the satellite dithering.} \\
\end{tabular}
\end{center}\end{sidewaystable*}

\section{Results}\label{sect:results}

\begin{figure*}
\begin{center}
\parbox{18cm}{
\parbox{6cm}{
\resizebox{6cm}{!}{\includegraphics{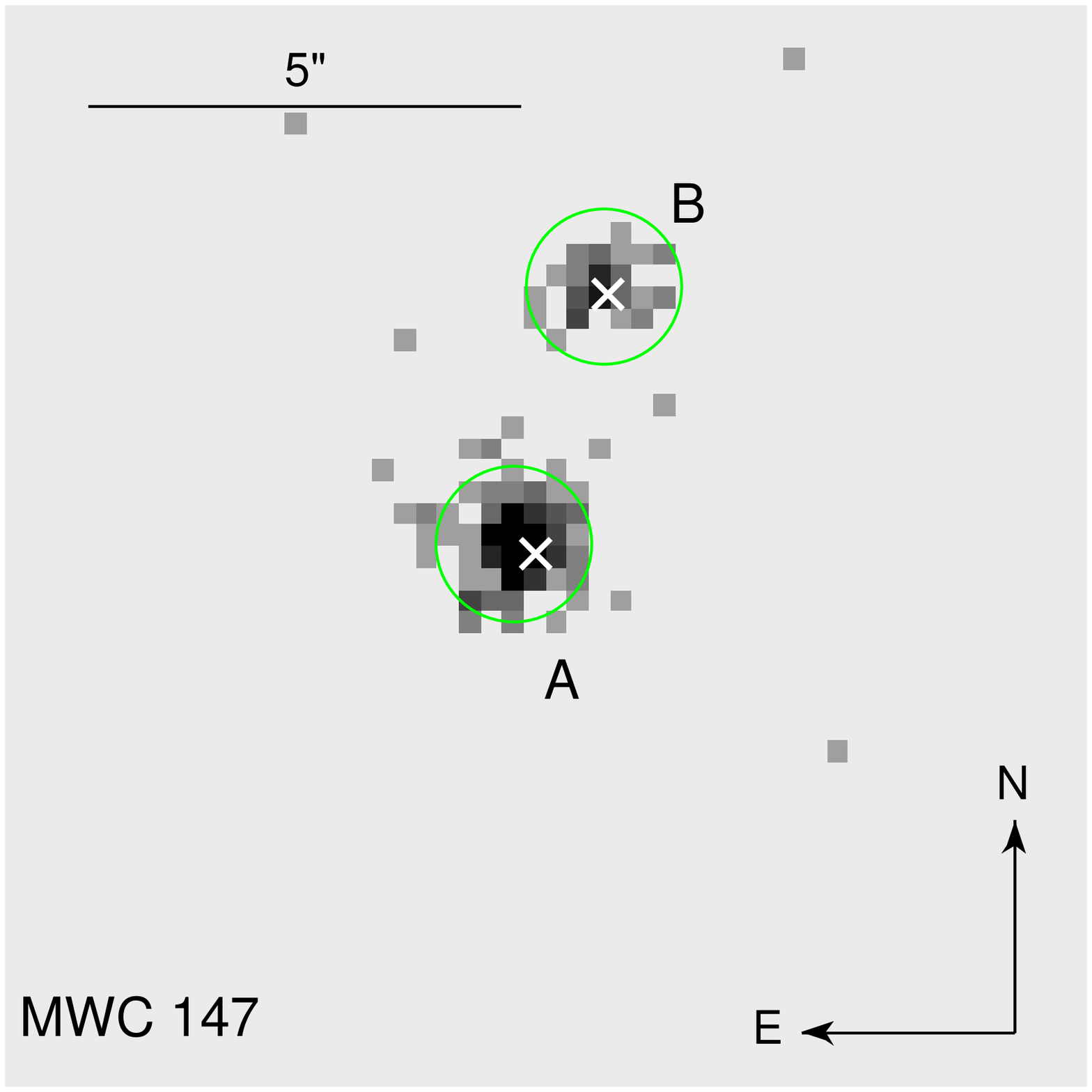}}
}
\parbox{6cm}{
\resizebox{6cm}{!}{\includegraphics{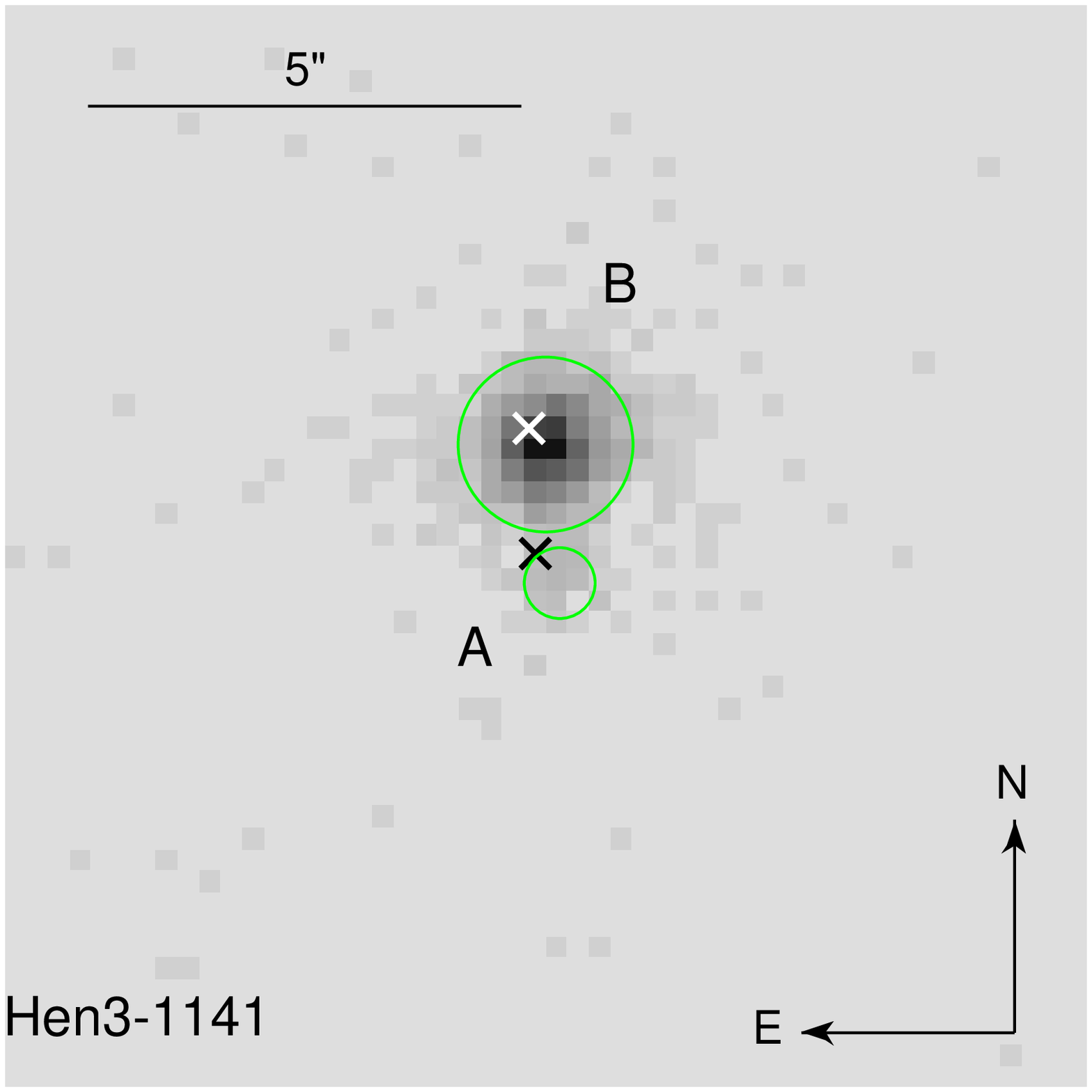}}
}
\parbox{6cm}{
\resizebox{6cm}{!}{\includegraphics{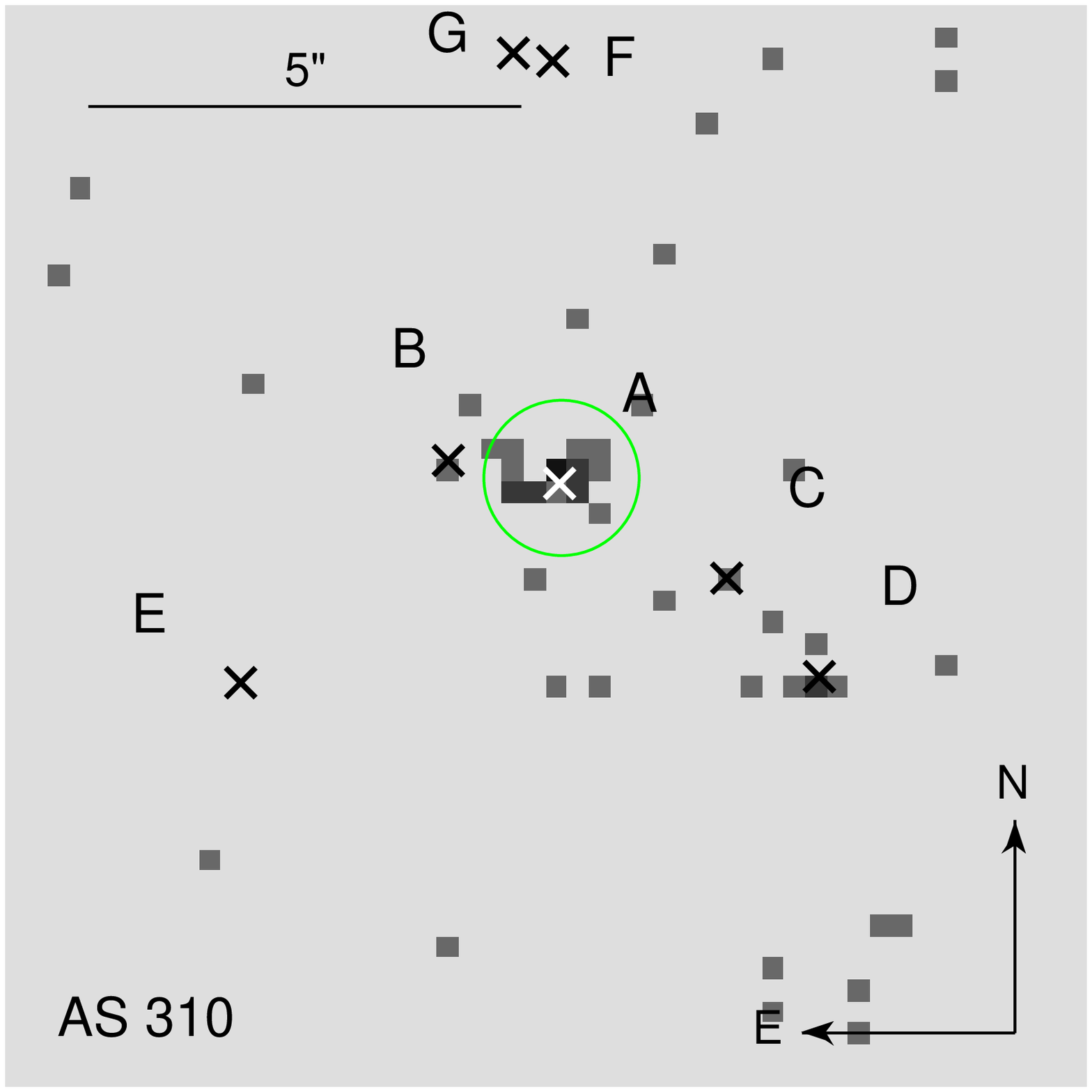}}
}
}
\parbox{18cm}{
\parbox{6cm}{
\resizebox{6cm}{!}{\includegraphics{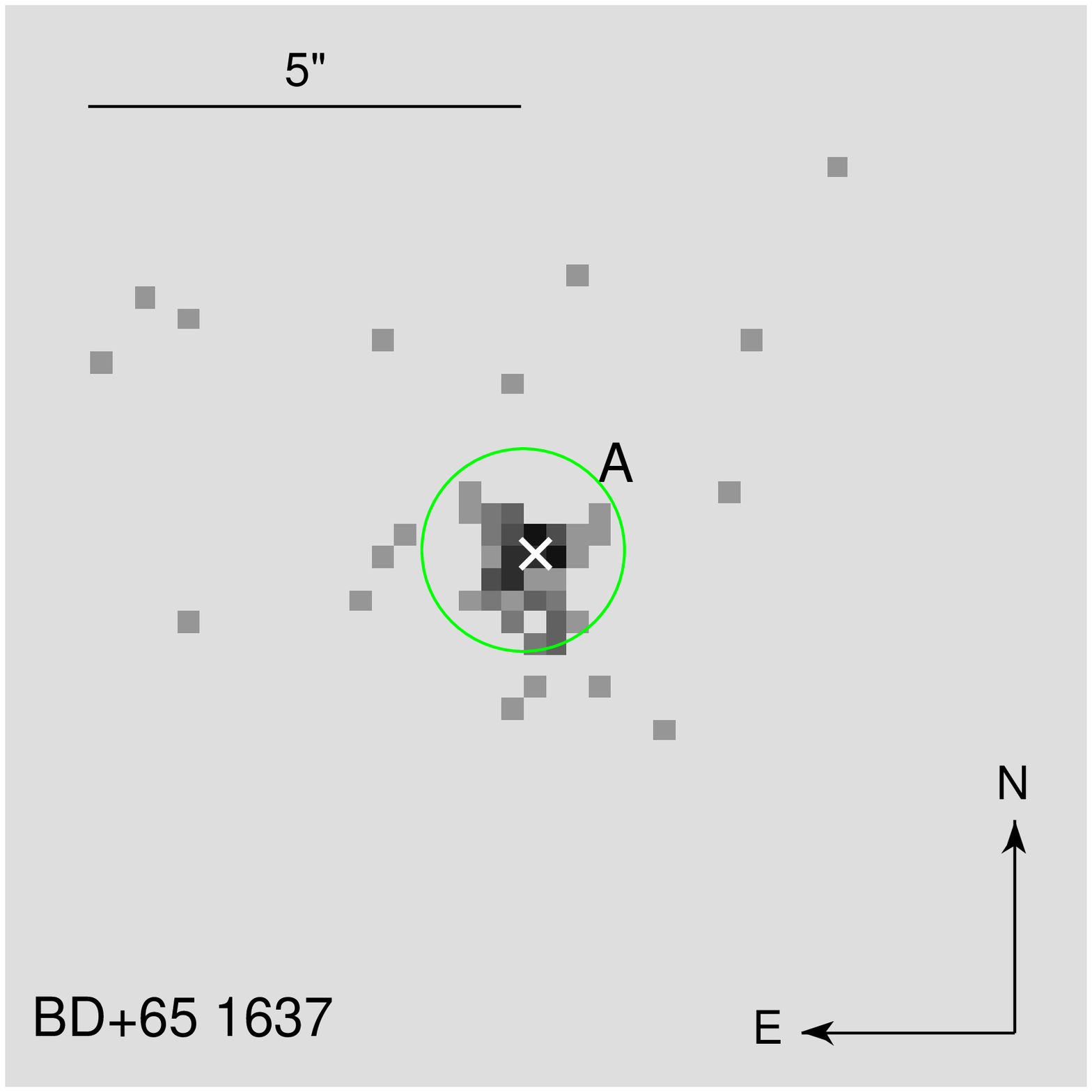}}
}
\parbox{6cm}{
\resizebox{6cm}{!}{\includegraphics{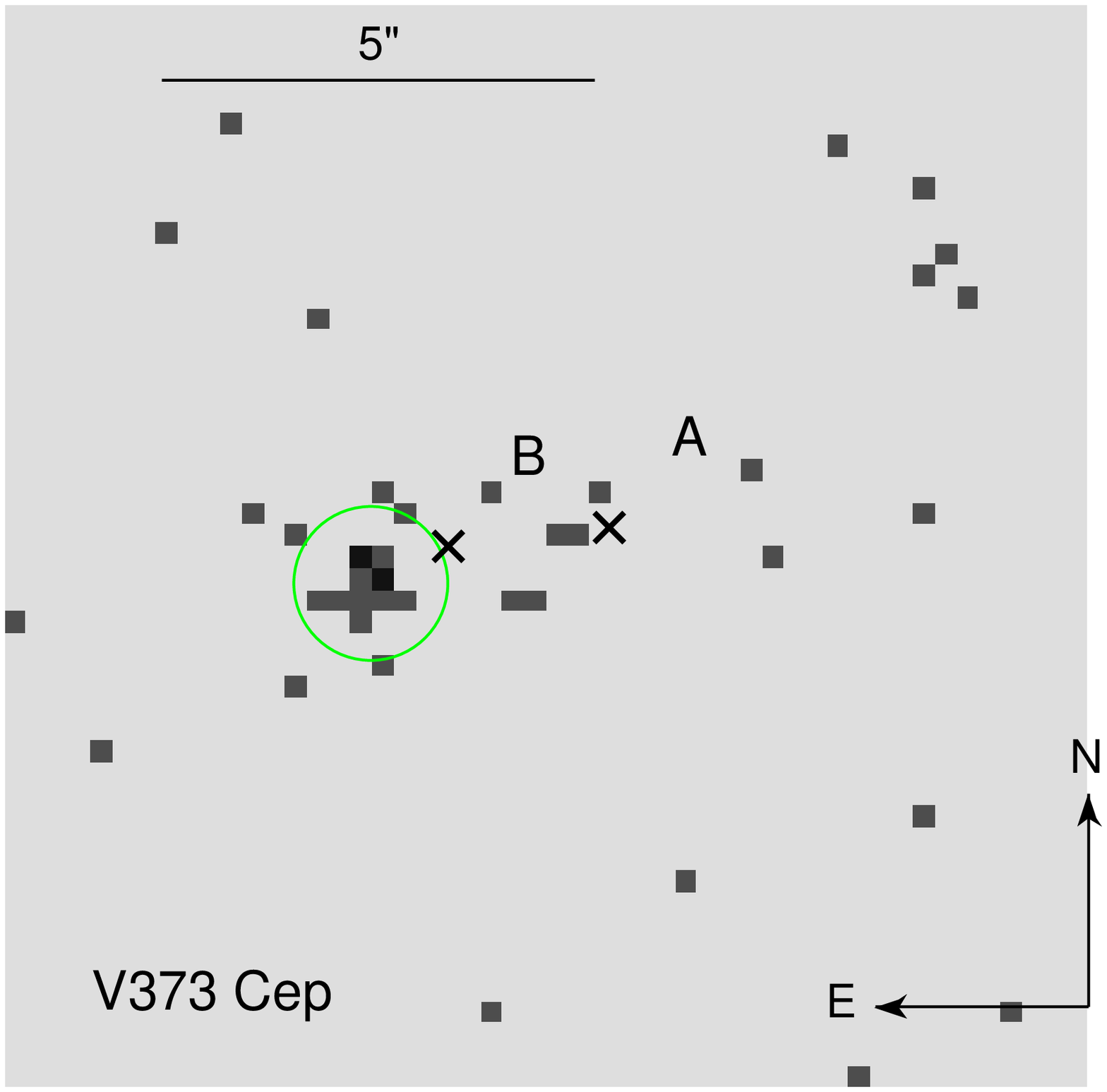}}
}
\parbox{6cm}{
\resizebox{6cm}{!}{\includegraphics{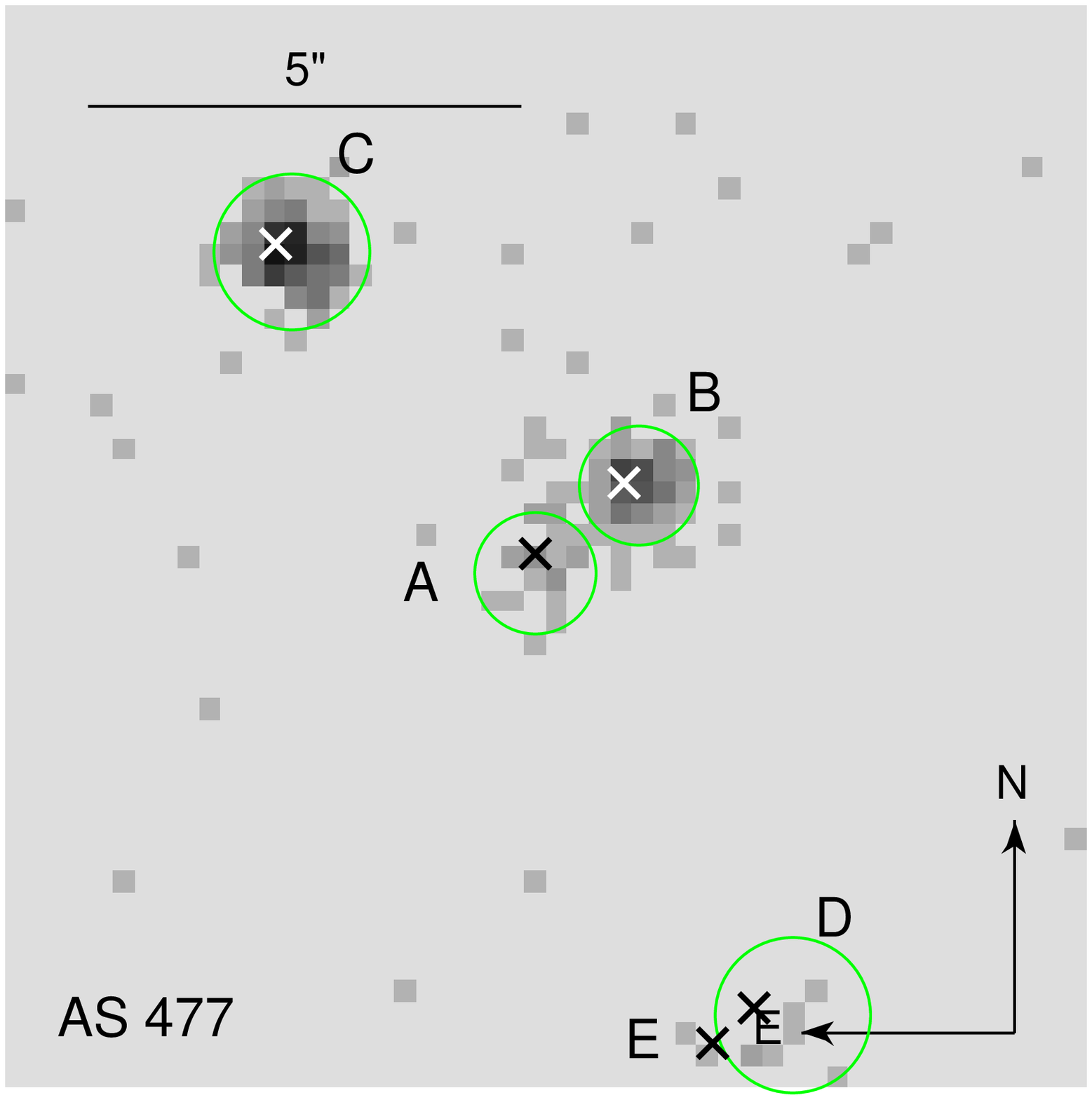}}
}
}
\parbox{18cm}{
\parbox{6cm}{
\resizebox{6cm}{!}{\includegraphics{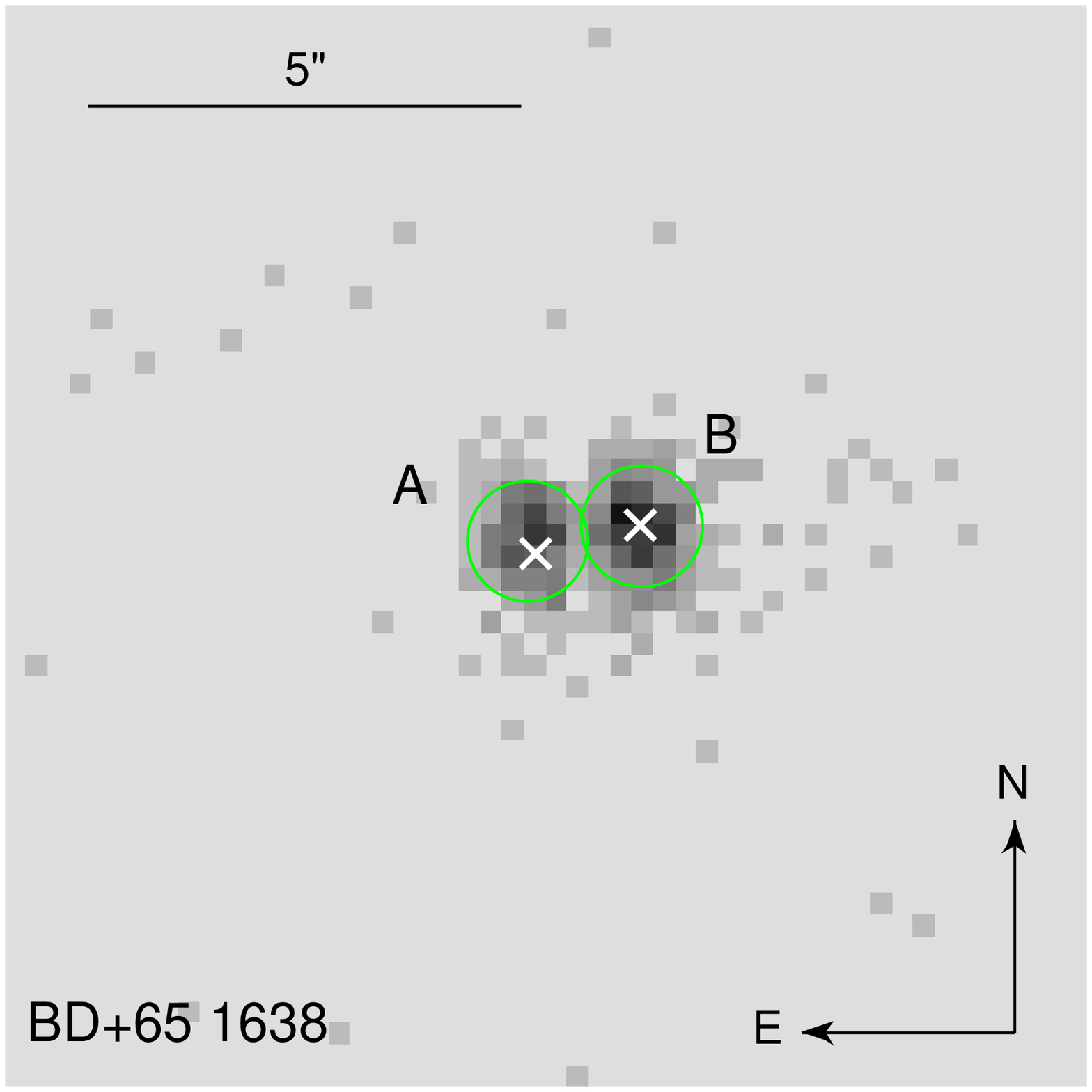}}
}
\parbox{6cm}{
\resizebox{6cm}{!}{\includegraphics{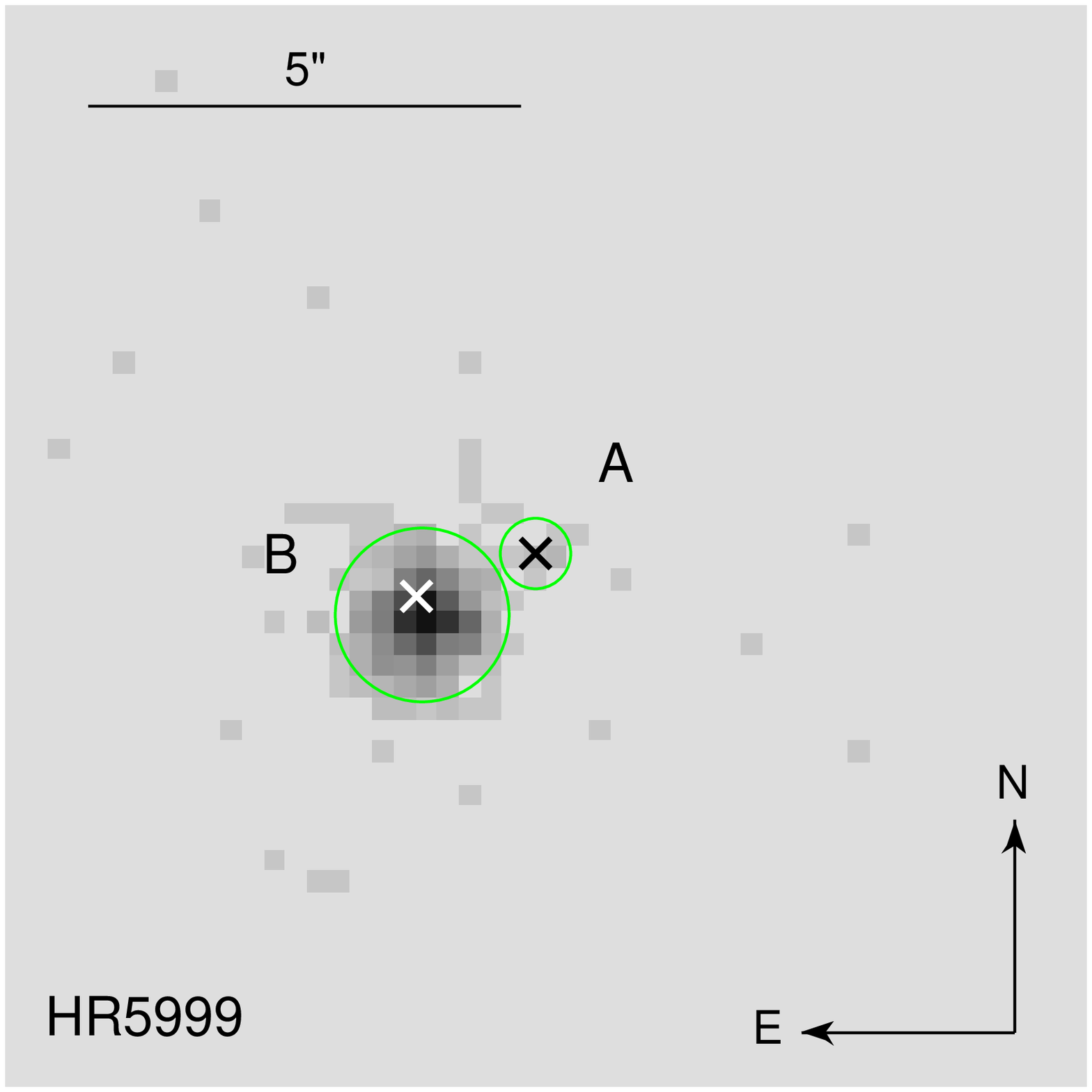}}
}
\parbox{6cm}{
\resizebox{6cm}{!}{\includegraphics{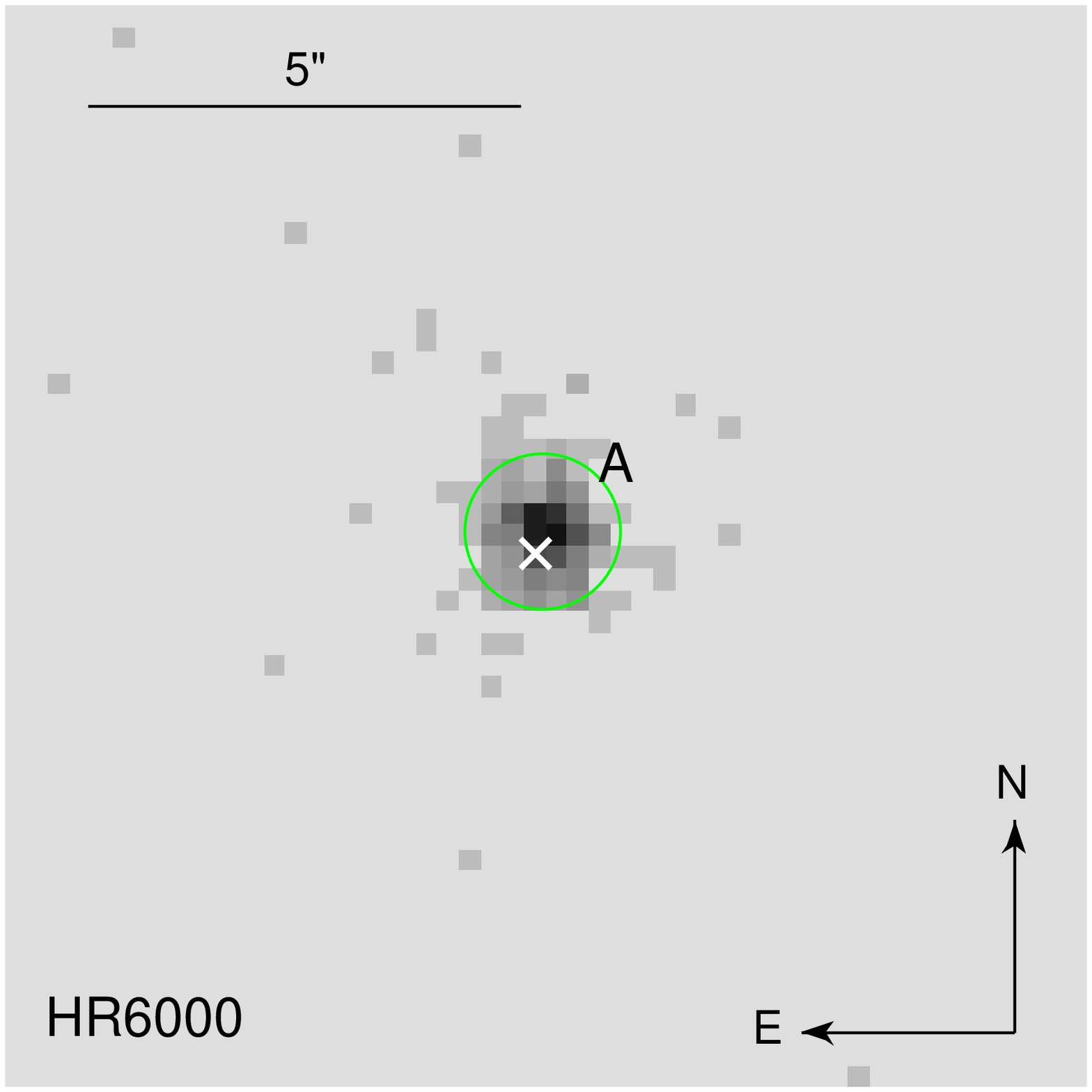}}
}
}
\caption{{\em Chandra} ACIS images of HAeBe stars binned to a pixel size of $0.25^{\prime\prime}$.  
Crosses denote the optical/IR position of the individual components in the multiple system, 
circles mark the photon extraction areas centered on the position of X-ray sources detected
 with {\sc WAVDETECT}.}
\label{fig:acis_images_haebe}
\end{center}
\end{figure*}

Fig.~\ref{fig:acis_images_haebe} shows the $25^{\prime\prime} \times 25^{\prime\prime}$ ACIS images 
centered on the primary stars. 
The photon extraction areas of all detected X-ray sources are overplotted (circles), 
as well as the position of the primary and the position of the companions (x-shaped symbols). 
The separation to each X-ray source detected in this image is measured for 
the optical/IR positions of the primary (HAeBe) stars and all known visual components.
When more than one of the optical components are identified with the same X-ray source
the X-ray detection is attributed to the closest one. In practice, this is relevant for 
cases where not all components are X-ray detected; see Fig.~\ref{fig:acis_images_haebe} (AS310, V373Cep). 

Table~\ref{tab:xrayparams_lx} summarizes the identification of all X-ray sources with components 
of our target systems and their X-ray parameters. 
Cols.~1-5 give
the designation of the target, 
component identifier, 
a flag for X-ray detections, 
offaxis angle, 
and offset between X-ray and optical/IR position. 
Since very faint sources may escape the automatic detection procedure described in 
Sect.~\ref{sect:observations}, we analysed the images independent of the {\sc wavdetect} results. 
Counts are extracted from a circular area centered on the optical/IR position of each visual 
component with photon extraction radius corresponding to $90$\,\% of the point-spread-function (PSF).
For sources which are not fully separated smaller non-overlapping radii were chosen 
(see col.~10 of Table~\ref{tab:xrayparams_lx}). 
We restrict the analysis to the $0.2-8$\,keV energy band. 
The background is measured in a squared area of 
$1^\prime$ side length centered on the optical/IR position of the respective
star but excluding all detected sources. After scaling to the source extraction area this
background is negligibly small in the chosen spectral range for all images ($< 1$\,cts).
The irrelevant influence of the background is also obvious from 
a look at the images in Fig.~\ref{fig:acis_images_haebe}. 
The number of net source counts ($N$) in the $0.2-8$\,keV passband is given
in column~7. Errors were computed with the Gehrels approximation $\sqrt{N+0.75} + 1$ for
Poisson distributed data \citep{Gehrels86.1}. 
Col.~6 represents the signal-to-noise ratio. 
Furtheron, we consider all stars with $S/N > 3$ detected. 
To compute upper limits for the undetected components of our target systems 
we used the method for Poisson-distributed counting data described by \citet{Kraft91.1}. 

Columns~$8$ and~$9$ of Table~\ref{tab:xrayparams_lx} show hardness ratios defined as 
$HR = (H-S)/(H+S)$, where $H$ and $S$ are the number of counts in a hard band
and in a soft band, respectively. 
$HR1$ is defined for the $0.2-1.0$\,keV ($S$) and the $1.0-5.0$\,keV ($H$) bands,
and $HR2$ for the $1.0-1.5$\,keV ($S$) and the $1.5-5.0$\,keV ($H$) bands.
Hardness ratios are evaluated only for those stars recovered in the automatic source detection
procedure. The remaining ones all have less than $10$ photons in total and are unsuitable
for a more detailed analysis. 

In col.~$11$ the PSF- and absorption-corrected X-ray luminosity
in the $0.2-8$\,keV band is given. 
The X-ray luminosities were computed
with PIMMS\footnote{The Portable Interactive Multi-Mission Simulator (PIMMS) is accessible at 
http://asc.harvard.edu/toolkit/pimms.jsp} 
assuming an iso-thermal emitting plasma with $kT=1$\,keV
and an absorbing column density $N_{\rm H}$ corresponding to the value derived from 
$A_{\rm V}$ according to the extinction law of \citet{Ryter96.1} 
($N_{\rm H}\,[10^{22}\,{\rm cm^{-2}}] = A_{\rm V}\,[{\rm mag}] \times 1.8 \times 10^{21}$). 
In our previous {\em Chandra} study of HAeBe stars we have shown that, albeit
the assumption of an iso-thermal plasma may not be appropriate, 
the X-ray luminosities derived from the above assumptions for the spectral parameters 
are in reasonable agreement with the values obtained from the actual X-ray spectrum \citep{Stelzer06.3}. 

The bolometric luminosities (col.~12) are known only for the primary stars. They represent the blackbody radiation from 
the stellar photosphere, without taking into account excess emission from circumstellar material seen at IR and radio
wavelengths; see Table~\ref{tab:targets}. 
A rough estimate using the flux ratios measured in the AO images
suggests bolometric luminosities for the companions that are at least $1-2$\,dex lower than those of the
primaries. Consequently, the $L_{\rm x}/L_{\rm bol}$ ratios of the companions are higher than those of the 
HAeBe stars by the same amount, i.e. on the order of $10^{-4}$ as is typical for low-mass pre-MS stars. 
We renounce on a more detailed analysis of the X-ray properties of the resolved companions as they are not relevant
for the search of the origin of the emission from the primary stars. 

Finally, col.$13$ of Table~\ref{tab:xrayparams_lx} represents the significance 
of variability according to a Kolmogorov-Smirnov (KS) test. 
Two stars are variable at $>95\,\%$ probability according to the KS test. However, from visual inspection
of the lightcurves it is evident that the variability
of HR\,6000 must be attributed to the satellite dithering, and the only truly variable object is AS\,477\,C. 
Given the generally short exposure times and faintness of the targets 
the lack of detectable variability in these observations is not surprising.

\section{Discussion}\label{sect:discussion}

Our high-resolution X-ray imaging study of $9$ young intermediate-mass stars comes up with the remarkable 
detection rate of $100$\,\%. This result confirms earlier findings of a large number
of X-ray emitters among HAeBe stars.
In our previous {\em Chandra} study of $17$ HAeBe stars we detected $13$ \citep{Stelzer06.3}. 
While at least $5$ of those are known SBs, only one of the $9$ new targets is
known to be an SB. In the selection of the new sample we have explicitly included stars
with negative results from spectroscopic binarity searches (see Table~\ref{tab:targets}). 

In \cite{Stelzer06.3} we found that the X-ray luminosities 
and temperatures of HAeBe stars are similar to those of their resolved visual late-type companions 
and also to young low-mass stars in the Orion Nebula Cluster (ONC). 
With respect to MS B-type stars they have on average hotter 
and more luminous X-ray emission. 
For all but three of the $9$ new targets analysed in this paper 
less than $100$\,photons were collected, owing to the large
distance of some objects. This makes the sample unsuited for a detailed spectral analysis.
We examine instead the hardness ratios. 

\begin{figure*}
\begin{center}
\parbox{18cm}{
\parbox{9cm}{
\resizebox{9cm}{!}{\includegraphics{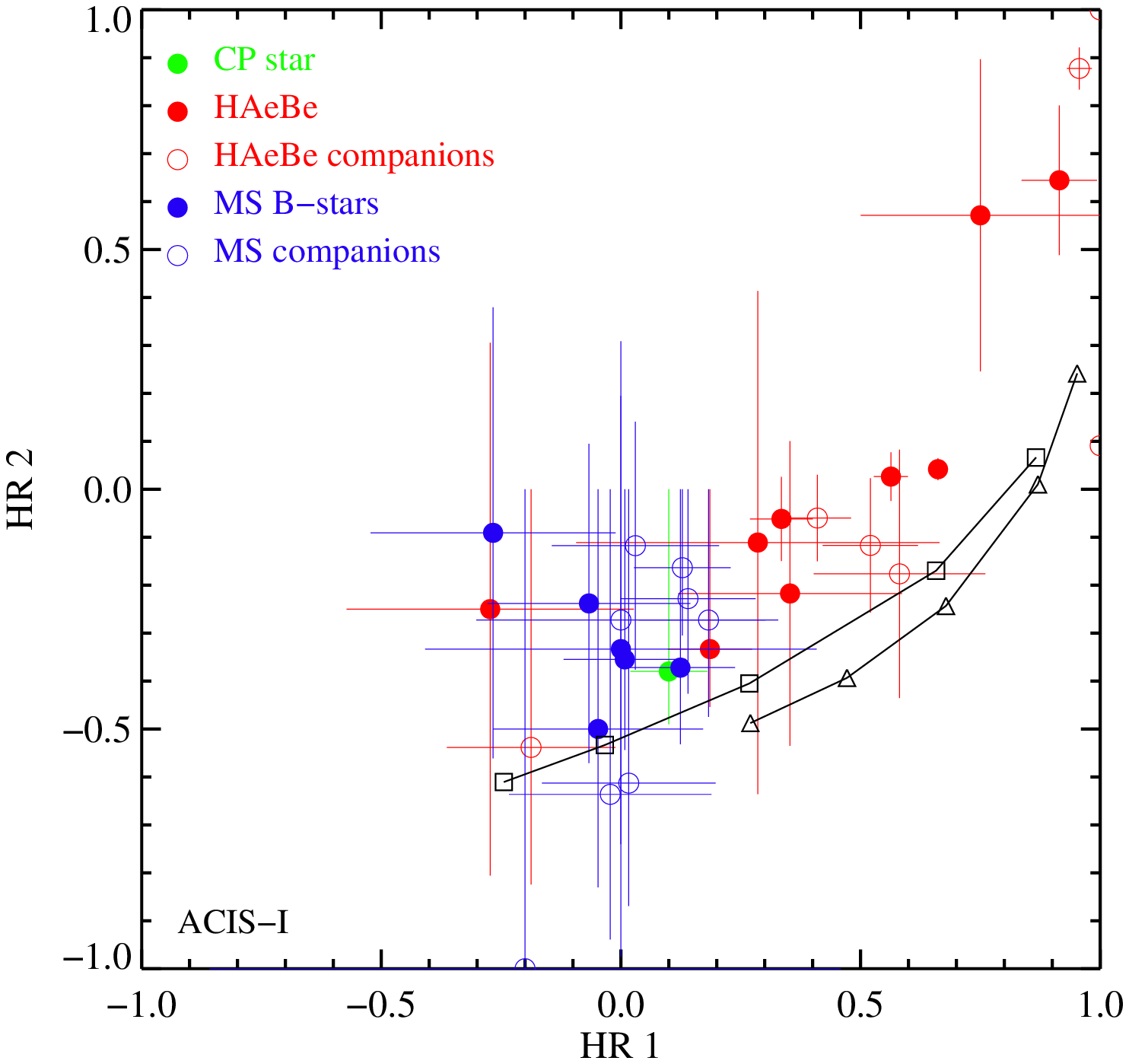}}
}
\parbox{9cm}{
\resizebox{9cm}{!}{\includegraphics{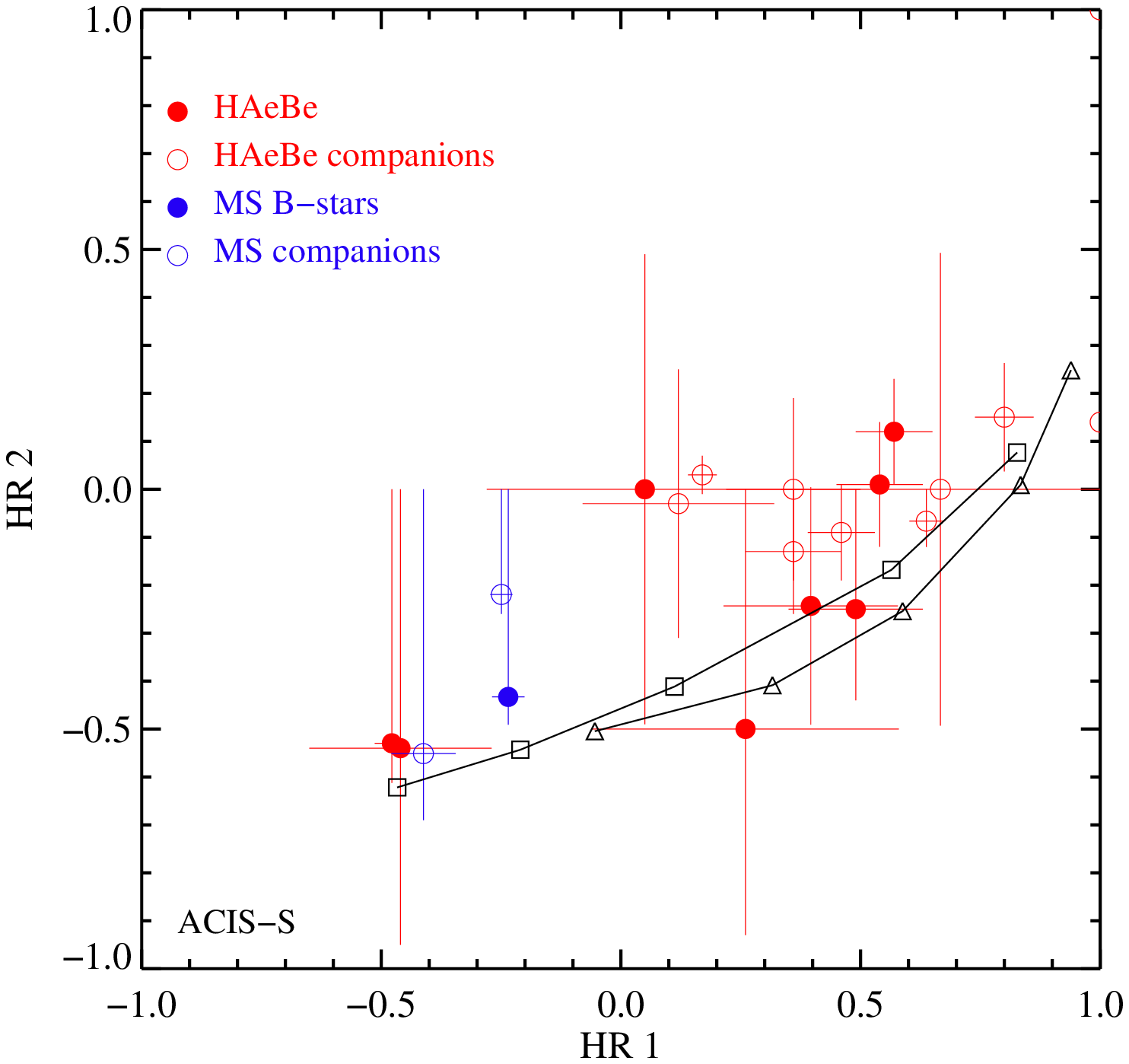}}
}
}
\caption{{\em Chandra} hardness ratios for Herbig stars, MS B-type stars and resolved visual companions. Data obtained
with the imaging and the spectroscopic array of ACIS are shown in separate graphs because of their different
spectral response. $HR1$ and
$HR2$ are defined as in Sect.~\ref{sect:results}. The solid lines represent calculated hardness ratios for the assumption
of an absorbed $1$-T thermal model with $kT = 0.7$\,keV (squares) and $kT = 1.0$\,keV (triangles). 
The symbols on those curves denote from left to right 
the hardness ratio for $N_{\rm H} = 0$, $0.2$, $0.5$, $1.0$, and $1.5\,10^{22}\,{\rm cm^{-2}}$.} 
\label{fig:hrs}
\end{center}
\end{figure*}

The boundaries for the energy bands defining the hardness
ratios defined in Sect.~\ref{sect:results} were empirically chosen such as to yield a wide spread
among the targets in the $HR2 - HR1$ plane. We compute hardness ratios in the same energy bands
for the stars presented previously by \cite{Stelzer03.1,Stelzer06.2,Stelzer06.3}, and display all
samples in Fig.~\ref{fig:hrs}. Observations obtained with ACIS-I and ACIS-S are considered separately.
The hardness ratios corresponding to an absorbed isothermal spectrum are also shown and explained in
the figure caption. 
The higher efficiency of ACIS-S for the detection of soft photons yields lower
values for $HR\,1$ with respect to ACIS-I for the same spectral shape.
The differences in the hardness ratios measured with ACIS-I and ACIS-S 
become negligible for strongly absorbed spectra due to the absence of soft photons.  
The Herbig stars, MS B-type stars, and (presumably late-type) companions 
are represented in different colors in Fig.~\ref{fig:hrs}. 
HR\,6000, which is a chemically peculiar and not a HAeBe star, is marked as green circle.
%
%

From the distribution of the hardness ratios 
no clear distinction is evident in Fig.~\ref{fig:hrs} between Herbig stars and their companion stars,
while the MS B-type stars have softer X-ray spectra. At face, this might indicate that the X-ray emitters
among the Herbig stars are additional unresolved late-type companions. Their X-ray luminosities 
($\log{L_{\rm x}}\,{\rm [erg/s]} = 28.3...31.2$) are also 
consistent with the typical range of T\,Tauri stars. 
However, as noted already by \cite{Stelzer06.3} and evident from  Fig.~\ref{fig:hrs} the companions of 
the MS B-type stars are also softer than the Herbig stars and their companions, suggesting an
evolutionary rather than a mass effect. 

It is impossible to assign an X-ray temperature $T_{\rm x}$ to the stars on basis of their hardness
ratios because $T_{\rm x}$ is degenerate with the column density $N_{\rm H}$: A given pair ($HR1$,$HR2$)
can indicate either low absorption and high temperature, or higher absorption combined with lower temperature. 
This is demonstrated in Fig.~\ref{fig:hrs} by the curves representing synthesized hardness ratios for
different values of ($T_{\rm x}$,$N_{\rm H}$). However, the softest spectra are evidently little absorbed,
and the hardest spectra indicate high column density, in line with the trend of decreasing absorption
along the evolutionary sequence. 
The fact that most data points lie above the
models shown in Fig.~\ref{fig:hrs} seems to indicate an excess in hard photons that can not be represented
by the simple $1$-T model assumed for calculating the expected colors. 

\begin{figure}
\begin{center}
\resizebox{9cm}{!}{\includegraphics{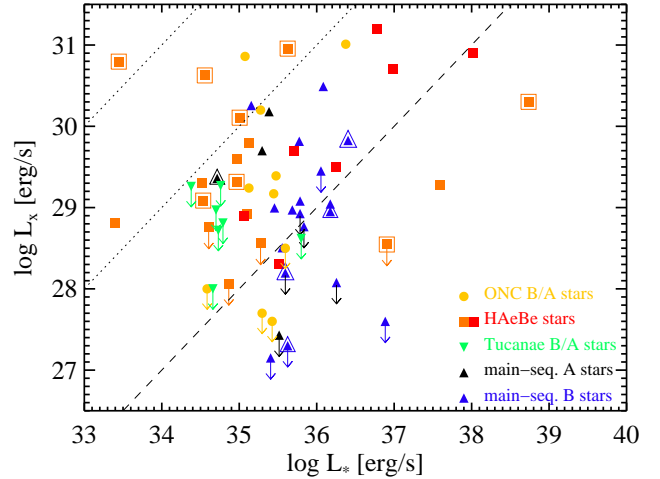}}
\caption{X-ray versus stellar luminosity for HAeBe stars and other samples of intermediate-mass stars (see text in Sect.~\ref{sect:discussion}). The new HAeBe targets and those from our previous works are distinguished by red and orange plotting symbols, respectively. Stars with unresolved companions are surrounded by larger symbols. The dashed line indicates a value of $L_{\rm x}/L_{\rm bol} = 10^{-7}$, the dotted lines represent levels of $10^{-5}$ and $10^{-3}$. 
}
\label{fig:lglx_lglbol}
\end{center}
\end{figure}

The $L_{\rm x}/L_{\rm bol}$ ratios for most HAeBe stars 
are between $\sim 10^{-5...-7}$ (Fig.~\ref{fig:lglx_lglbol}) confirming historic results. 
It strikes that all HAeBe stars from our studies 
for which $L_{\rm x}/L_{\rm bol}$ levels above $10^{-5}$ are measured 
have been identified as binaries that remain unresolved even with {\em Chandra}. 
From Fig.~\ref{fig:lglx_lglbol} we infer a possible splitting of the HAeBe sample in 'classical'
HAeBe systems of spectral type late-B/A and early-B type pre-MS stars:  
The true X-ray emission level of late-B/A Herbig stars -- if any -- seems to be on the order of
$\log{L_{\rm x}}\,{\rm[erg/s]} \sim 28...30$. The X-ray production efficiency of 
these stars is reduced with respect to T\,Tauri stars, which are generally characterized
by $L_{\rm x}/L_{\rm bol} \sim 10^{-3...-5}$. 
In Fig.~\ref{fig:lglx_lglbol} we include for comparison 
intermediate-mass stars of the $1$\,Myr-old ONC \citep{Stelzer05.1}, 
of the $\sim 30$\,Myr-old Tucanae association \citep{Stelzer00.2}, 
and of the field population \citep{Stelzer03.1, Stelzer06.2, Czesla07.1}.  
About half of the B- and A-type MS stars are found in the same range of $L_{\rm x}/L_{\rm bol}$ 
as the late-B/A HAeBe stars, 
while the others are located at levels well below $L_{\rm x}/L_{\rm bol} = 10^{-7}$
or altogether X-ray dark.
This points either at a population of unresolved late-type sub-arcsecond companions to late-B/A 
HAeBe stars or to a fading of the X-ray emission from late-B/A stars with age.  

To satisfy the latter hypothesis, a mechanism that works only at young ages is required. 
Star-disk magnetic interactions are unlikely to explain all of the X-ray detections among HAeBe stars,
as this mechanism is equally valid for low-mass T Tauri stars but not observed to be the major X-ray
production process. 
As outlined in Sect.~\ref{sect:intro}, 
possible mechanisms for intrinsic X-ray production from HAeBe stars include a fossil 
magnetic field, MCWS and shear dynamos, all of which require the presence of magnetic fields. 
Two of the eight HAeBe stars with positively detected magnetic fields have been observed in X-rays at
high spatial resolution with {\em Chandra}. These two stars are late-B/A HAeBe's and 
they are X-ray sources. 
However, the more relevant number 
for the understanding of the origin of their X-ray emission is how many X-ray detected
HAeBe stars have a null result from magnetic field studies. This number is not easily accessible from the 
literature because in some publications only the stars with positive magnetic field detections 
are listed and not the whole observed sample; e.g. \cite{Wade05.1} and \cite{Hubrig04.3}.
For some X-ray detected HAeBe stars from the Chandra sample no magnetic field could be detected with
spectropolarimetric methods \citep{Wade07.1}. 

In any case, the fraction of (apparently) X-ray emitting 
HAeBe stars seems to be much larger than the fraction of (apparently) magnetic HAeBe stars.
This questions any field related emission mechanism for the bulk of HAeBe stars 
in favor of the companion hypothesis, although magnetic fields may play a crucial role in explaining
the X-ray emission from some individual objects. As described in Sect.~\ref{sect:intro}, the case of
AB\,Aur provided convincing evidence for X-ray emission from the wind region and magnetic fields are
needed to explain its X-ray variability \citep{Telleschi07.2}.
For the A1e star HD\,163296 the unusually soft X-ray emission was explained by emission
from an accretion shock at the end points of magnetospheric funnel flows 
similar to the case of some lower-mass cTTS \citep{Swartz05.1}. However, recent analysis suggests that its 
X-ray properties may resemble those of AB\,Aur: cool and hot
X-ray emitting plasma but no evidence for high densities as seen in cTTS making accretion
shocks as origin of the X-ray production unlikely and suggesting an emission site above the
surface of the star (Guenther et al., in prep.).  
Interestingly, the polarization signatures in some HAeBe stars show evidence for a origin in 
circumstellar matter rather than on the stellar surface suggesting magnetic fields associated
with the wind \citep{Hubrig07.1}. 

The only HAeBe stars in our high-resolution {\em Chandra} studies that showed high X-ray luminosities
($\log{L_{\rm x}}\,{\rm [erg/s]} \sim 31$) and no evidence for unresolved companions are of early spectral type  
(cf. Fig.~\ref{fig:lglx_lglbol}). If considered intrinsic to the HAeBe stars their X-ray emission scales
with bolometric luminosity similar to hot stars ($\log{L_{\rm x}/L_{\rm bol}} \sim -6...-7$).
Nevertheless, it can not be excluded that the bulk of the emission of early-B HAeBe stars 
comes from unknown cooler companions. 

The six HAeBe stars selected for the study presented in this paper have known visual companions 
resolved with {\em Chandra}, while no reports on binarity are known for the three B(e) stars
added to the sample because they were serendipitously observed in the same {\em Chandra} fields. 
This implies that, if the X-rays detected in this survey are to be ascribed 
to cool companion stars all of the six original targets are triple systems. 
Little is known about the higher order multiplicity 
of HAeBe stars. In any case, 
high-resolution imaging studies came up with an enhanced binary frequency of Herbig stars 
when compared to solar-type MS stars \citep[e.g.][]{Leinert97.1, Bouvier01.1}. 
This is even more evident in a recent spectro-astrometric search for sub-arcsecond companions 
to HAeBe stars where \cite{Baines06.1} reported with $68$\,\% the highest multiplicity fraction observed so far
in any sample of these type of stars.

\section{Conclusions}\label{sect:conclusions}

In summary, this study remains inconclusive as to the origin of the X-ray sources.  
Either HAeBe stars are intrinsic hard X-ray sources with similar
properties as lower-mass T Tauri stars but scaled down with respect to the stellar luminosity
or the majority of the low-mass companions of the HAeBe stars has not yet been identified. 
From a consideration of the X-ray luminosities 
we have speculated about a possible division in two types of HAeBe stars. The hotter group of 
early-B stars might behave similar to early-type stars on the MS, and the cooler group of 
'classical' late-B/A HAeBe stars might involve magnetic field related X-ray emission processes 
as had been suggested for the prototype of this class, AB\,Aur. 
First indications for different X-ray brightness of Herbig Ae and Be stars had been mentioned by
\cite{Damiani94.2}, however on basis of a sample observed by low spatial resolution with {\em Einstein} and 
dominated by upper limits. As a result of our high-resolution, high-sensitivity 
{\em Chandra} imaging several HAeBe stars have now been resolved from late-type companion stars. 
There is evidence that the intrinsic X-ray emission level of late-B/A Herbig stars 
-- if any -- is $\sim 10^{29}$\,erg/s,
one to two orders of magnitude lower than derived from earlier low-resolution X-ray studies. 
A more detailed analysis for the viability of the potential emission processes in our targets is not feasible, 
owing to the insufficient diagnostic power of the existing X-ray observations, 
the absence or inhomogeneity of the database for wind properties, and 
a lack of quantitative predictions of some theoretical models.  
As demonstrated for the case of AB\,Aur, 
useful constraints can be placed by means of high-resolution X-ray spectroscopy
that provides access to source density and location and by means of X-ray monitoring 
searching for variability.
Furthermore, a comprehensive multiplicity study for intermediate-mass 
stars in various evolutionary stages is needed and is underway (Thomas et al., in prep.). 

%
%

\appendix

\section{Individual Systems}\label{sect:indiv}

\subsection{MWC\,147}\label{subsect:mwc147}

MWC\,147 (HD\,259531) is a well-studied Herbig Be star in Monoceros \cite[$800$\,pc; ][]{Herbst82.1}.
The {\em Hipparos} parallax gives a smaller distance of 290pc \citep{Bertout99.1}. 
MWC\,147 is strongly accreting. Mass accretion rates from 
$\dot{M}_{\rm acc} \sim 10^{-5}$ to $10^{-7}\,{\rm M_\odot/yr}$
have been estimated with different diagnostics including radio emission, 
near-IR line emission and SED fitting \cite[see][for a summary]{Kraus08.1}. 
A search for magnetic fields did not result in a significant detection \citep{Wade07.1}.
\cite{Mottram07.1} interpreted the observed line depolarization as evidence for accretion 
from the disk without the mediation by a magnetosphere. 
 
MWC\,147 has a faint visual companion at $3.1^{\prime\prime}$ separation \citep{Baines06.1},
and no evidence for being a spectroscopic binary \citep{Corporon99.1}. However, \cite{Akeson00.1} found that 
IR interferometric visibility data 
of MWC\,147 cannot be described by a standard accretion disk model but a good fit
is obtained if an additional companion at milli-arcsecond separation is assumed. 

Our high spatial resolution image with {\em Chandra} has shown that the 
earlier {\em Einstein} X-ray source reported by \cite{Damiani94.2} 
includes contributions from both MWC\,147 
and its companion. The X-ray luminosity measured with {\em Einstein} 
is a factor two lower than the combined X-ray luminosity of MWC\,147\,A+B obtained
with {\em Chandra}, part of which can be explained by different assumptions on the
column density. MWC\,147 has the highest X-ray luminosity in this sample. This might
be an indication for shorter distance as suggested by the {\em Hipparcos} measurement.

\subsection{Hen\,3-1141}\label{subsect:hen3_1141}

Hen\,3-1141 is a late-A or early-F star, one of the latest in spectral type among the HAeBe 
stars in Th\`e's catalog.
In the past it had been erroneously classified as a post-AGB star \citep{Oudmaijer92.1} or a 
Vega-type star \citep{Walker88.1}. \cite{Perez04.1} detected Lithium absorption and 
P\,Cyg profiles in the H$\alpha$ emission evidencing 
the presence of winds and thus supporting the pre-MS character of Hen\,3-1141. 
The sky position of Hen\,3-1141 suggests membership in Sco\,OB2 ($\sim 145$\,pc, although 
a larger distance of $250$\,pc was inferred from the {\em Hipparcos} parallax.
A weak magnetic field was marginally ($1.6\sigma$) detected by \cite{Hubrig04.3}. 

A faint object located $1.4^{\prime\prime}$ to the north of Hen\,3-1141 detected on
near-IR images was confirmed to share proper motion with Hen\,3-1141. 
This companion was classified as a K2 star based on the strength of its Na\,I 
doublet. \cite{Perez04.1} argued that the {\em Hipparcos} measurement might be
influenced by binary motion. For a distance of $\sim 145$\,pc the 
two components of the binary star are coeval on evolutionary tracks, and we follow
the suggestion of \cite{Perez04.1} for an association of Hen\,3-1141 with Sco\,OB2. 

No previous reports on X-ray emission are found in the literature. With {\em Chandra}
we detected X-rays from both the Herbig star and its northern companion.

\subsection{AS\,310}\label{subsect:as310}

AS\,310 is associated with the H\,II region S\,61. Long known to be a $4.4^{\prime\prime}$ 
binary \citep{Bastian79.1}, \cite{Ageorges97.1} discovered four further near-IR sources within 
$5^{\prime\prime}$ of the HAeBe star, probably representing the brightest members of a young
star cluster \citep{Testi98.1}. One of these objects was resolved in a double source by
Thomas et al., in prep. \cite{Polomski02.1} detected an extended mid-IR nebulosity coincident 
with these companions. Source\,E of Thomas et al., in prep. is also reported by 
\cite{Maheswar02.1}, and sources\,E and F by \cite{Polomski02.1}. 

No previous reports on X-ray emission are found in the literature. The brightest X-ray emission
observed with {\em Chandra} in the area is associated with the Herbig star, 
and some of the companions are weakly detected.

\subsection{HAeBe stars in NGC\,7129}\label{subsect:v373cep}

The NGC\,7129 reflection nebula at a distance of $1000$\,pc \citep{Hillenbrand92.1, Finkenzeller84.1} 
is illuminated by the B-type stars BD\,+65\,1637 and BD\,+65\,1638 that have evacuated their environment
of molecular material and are surrounded by a cluster of low-mass stars. The star cluster was recently
studied with {\em Spitzer} \citep{Gutermuth04.1, Muzerolle04.2}. 
A third Herbig Be star, V373\,Cep (better known as LkH$\alpha$\,234), is located on a molecular ridge at the
eastern end of the cluster. LkH$\alpha$\,234 is the youngest of the HAeBe stars in NGC\,7129. 

An optically thick disk around LkH$\alpha$\,234 was inferred from the slope of its spectral energy 
distribution \citep{Fuente01.1}. \cite{Chakraborty04.1} observed correlated variability of H$\alpha$ 
and He\,I that is considered as indicator of ongoing accretion \citep{deWinter99.1}. 
Transient redshifted absorption components in Na\,D lines, uncorrelated with the
changes in H$\alpha$ and He\,I lines, were explained with an infalling planetesimal. In this scenario the 
dust disk is in a relatively evolved stage.  
Indeed, the outflows detected near LkH$\alpha$\,234 could be associated to a nearby deeply embedded 
infrared companion, IRS\,6,  located $\sim 4^{\prime\prime}$ NW of the Herbig star \cite{Fuente01.1}. 
In the same vein, none of the five radio emitters within $5^{\prime\prime}$ of LkH$\alpha$\,234,  
that are interpreted as thermal radio jets and CO outflows,  
is powered by the Herbig Be star suggesting it is more evolved and less embedded \citep{Trinidad04.1}. 
Several IR sources were reported in the vicinity of V373\,Cep \citep{Weintraub94.1}. 
None of these radio and IR emitters is detected with {\em Chandra}. 

Previous X-ray surveys had provided a confusing picture of the region. 
The {\em Einstein} IPC could not resolved LkH$\alpha$ and BD\,+65\,1637. A strong but largely
displaced X-ray source ($\log{L_{\rm x}}\,{\rm [erg/s]} = 31.7$; $0.9^\prime$) was assigned to both HAeBe stars
\citep{Damiani94.2}. No X-ray emission was detected with {\em ROSAT} yielding an upper limit of 
$\log{L_{\rm x}}\,{\rm [erg/s]} < 30.7$ for LkH$\alpha$\,234, higher than and therefore consistent
with the new {\em Chandra} detection. The only {\em ROSAT} X-ray source in the region was 
associated with the third Herbig Be star of NGC\,7129, BD+65\,1638 \citep{Zinnecker94.1}. 
With {\em Chandra} BD+65\,1638 is resolved into a double X-ray source possibly
indicating binarity. Both sources are nearly equally bright in X-rays. The stellar parameters
of BD+65\,1638 are not known and this system deserves follow up.

\subsection{AS\,477}\label{subsect:as477}

AS\,477, better known as BD\,+46\,3471, 
has evacuated a cavity in the molecular cloud and was classified a star 
that has dispersed its circumstellar material \citep{Fuente02.1}. The existence
of outflow activity is manifest in [OI]\,$6300$\,\AA~emission \citep{Corcoran98.1}, 
and a wind with moderate mass loss rate of $\sim 10^{-7}\,{\rm M_\odot/yr}$ was diagnosed
by modeling of emission lines \citep{Nisini95.1, Bouret98.1}. 
 
Thomas et al., in prep. find four faint IR sources within $6^{\prime\prime}$ of AS\,477.
Component C was earlier reported by \cite{Pirzkal97.1} and by 
\cite{Maheswar02.1}. 
There is no evidence for a spectroscopic companion \citep{Corporon99.1}. 

Although AS\,477 is known to be surrounded by further emission line stars (LkH$\alpha$235...239 
within $1^\prime$; Herbig 1960) \cite{Zinnecker94.1} assigned the strong {\em ROSAT} 
X-ray source ($\log{L_{\rm x}}\,[{\rm erg/s}] = 31.2$) to AS\,477. {\em Chandra} has now
shown that, although the Herbig star is detected, 
most of the X-ray emission is produced by the two closest companion stars.  

\subsection{HR\,5999\,/\,HR\,6000}\label{subsect:hr5999}

The stellar pair HR\,5999/HR\,6000 is located in the central part of the Lupus~3 dark cloud at a 
distance of about 150\,pc \citep[see e.g.][]{Hughes93.1}.
HR~5999 (=V856 Sco) with spectral type A7IVe belongs to the class of HAeBe stars
\citep{The94.1} showing all the required phenomenology
such as photometric variability and emission lines. Strong evidence for a
disk exists from the IR energy distribution \citep{Hillenbrand92.1}. 
\cite{Stecklum95.1} report IR astrometry of a fainter companion around HR~5999 
(a 0.5~M$_{\odot}$ star named Rossiter~3930) at a distance of $1.5^{\prime\prime}$.  

HR\,5999 forms a common proper motion pair with HR\,6000 (=V1027 Sco, A1.5III) 
at $45^{\prime\prime}$, corresponding 
to a projected separation of $6300$\,AU when adopting a distance of $150$\,pc. 
We note that the {\it Hipparcos} distances are slightly larger ($208$ and $241$\,pc for HR\,5999 and HR\,6000,
respectively).
From optical spectroscopy \citet{Andersen84.1} concluded that HR\,6000 is a single, 
slowly rotating (or pole on), chemically peculiar 
He-weak late-B/early A-type star without any evidence for circumstellar matter.  As such it would be an
interesting candidate for an intrinsic A-type X-ray emitter, since chemically peculiar stars often do have 
magnetic fields. 
However, \citet{vandenAncker96.1} interpret its peculiar spectral and photometric properties as evidence 
for HR\,6000 being a young binary system containing a B6V star and
a cool T~Tauri companion. This scenario would also explain the weak IR excess and the X-ray 
emission detected from HR\,6000.

X-ray emission from HR\,5999 and HR\,6000 was already detected with the {\it ROSAT} PSPC 
\citep{Zinnecker94.1}. 
\citet{Hamaguchi05.1} report an {\it ASCA} detection of HR\,5999 and both stars were also 
observed by {\it XMM-Newton}. We investigated this archival {\it XMM-Newton} observation
and find that the X-ray emission from HR\,5999 is more strongly absorbed 
than that from HR\,6000, in accordance with the optical extinction.
Since the light paths from HR\,5999 an HR\,6000 have very similar trajectories,
the absorption observed towards HR\,5999 must then occur in the vicinity of the star itself. 
All previous X-ray observations suffered from low spatial resolution. 
In the {\em Chandra} data presented here, HR\,5999 is for the first time resolved from its late-type
companion, which turns out to be the brighter X-ray source although weak X-ray emission is also detected
from the primary Herbig star. 
HR\,6000 can not be resolved with {\em Chandra} from its spectroscopic companion, and the origin of
its X-ray emission remains unclear.

\begin{acknowledgements}
We would like to thank the referee, M.\,G\"udel, for constructive comments.
B.S. acknowledges financial support from ASI-INAF\,I/088/06/0 and J.R. from DLR under 50QR0803.
This research has made use of data obtained with {\it Chandra} under proposal numbers 07200005 and 09200162, 
software provided by the Chandra X-ray Center (CXC),
the SIMBAD database (operated at CDS, Strasbourg, France),
and the {\em Hipparcos} catalogue accessed through the VizieR data base. 
\end{acknowledgements}

\bibliographystyle{aa} 
\bibliography{stelzer}

\end{document}